\documentclass[12pt]{article}
\usepackage{mathrsfs}
\usepackage{amssymb}
\usepackage{amsmath}
\usepackage[amsmath,thmmarks]{ntheorem}
\usepackage{cite}

\usepackage{graphicx}
\usepackage{dcolumn}
\usepackage{bm}
\usepackage{slashed}

\newcommand{\be}{\begin{eqnarray}}
\newcommand{\ee}{\end{eqnarray}}
\newcommand{\ra}{\rightarrow}

\begin{document}

\begin{titlepage}

\makebox[6.5in][r]{\hfill ANL-HEP-PR-10-60}

\vskip1.40cm
\begin{center}
{\Large {\bf FEWZ 2.0: A code for hadronic Z production at next-to-next-to-leading order}} \vskip.5cm
\end{center}
\vskip0.2cm

\begin{center}
{\bf Ryan Gavin$^1$, Ye Li$^1$, Frank Petriello$^{2,3}$, and Seth Quackenbush$^2$}
\end{center}
\vskip 6pt
\begin{center}
{\it $^1$Department of Physics, University of Wisconsin, Madison, WI 53706, USA} \\
{\it $^2$High Energy Physics Division, Argonne National Laboratory, Argonne, IL 60439, USA} \\
{\it $^3$Department of Physics \& Astronomy, Northwestern University, Evanston, IL 60208, USA} \\
\end{center}

\vglue 0.3truecm

\begin{abstract}
\vskip 3pt \noindent
We introduce an improved version of the simulation code FEWZ ({\bf F}ully {\bf E}xclusive ${\bf W}$ and ${\bf Z}$ Production) for hadron collider production of 
lepton pairs through the Drell-Yan process at next-to-next-to-leading-order (NNLO) in the strong coupling constant.  The program is fully 
differential in the phase space of leptons and additional hadronic radiation.  The new version offers users significantly more options for customization.  
FEWZ now bins multiple, user-selectable histograms during a single run, and produces parton distribution function (PDF) errors automatically.  It also features 
a signifcantly improved integration routine, and can take advantage of multiple processor cores locally or on the 
\emph{Condor} distributed computing system.  We illustrate the new features of FEWZ by presenting numerous phenomenological results for LHC physics.  We 
compare NNLO 
QCD with initial ATLAS and CMS results, and discuss in detail the effects of detector acceptance on the measurement of angular quantities associated with 
$Z$-boson production.  
We address the issue of technical precision in the presence of severe phase-space cuts.

\end{abstract}

\end{titlepage}
\newpage

\section{Introduction \label{intro}}
Electroweak gauge boson production is a standard candle for Large Hadron Collider (LHC) physics studies.  It is a background to $Z'$ 
production and numerous other new physics searches.  As one of the cleanest processes with copious production (millions per year at 
design luminosity), it can be used as a luminosity monitor~\cite{Dittmar:1997md}, to constrain PDFs (parton distribution functions)~\cite{pdf}, and to 
study electroweak physics 
parameters~\cite{Haywood:1999qg}.  Therefore, understanding its production is crucial as the LHC physics program moves forward.  With so many events, 
systematic errors dominate statistical ones, and are expected to eventually reach $1-2\%$~\cite{systematics} 
at the LHC.  Next-to-leading order (NLO) predictions in the strong coupling, with $O(10\%)$ errors, are insufficient for a precise comparison with data; 
more accurate calculations are required.  Predictions through NNLO in perturbative QCD must be used.

The inclusive $O(\alpha_S^2)$ corrections to electroweak gauge boson production have been known for some time~\cite{Hamberg:1990np}. The theoretical 
uncertainties are at the percent level.   Exclusive production, which is necessary for any realistic prediction or phenomenological study in a detector 
of finite acceptance, is technically challenging but has been achieved~\cite{Anastasiou:2003yy,Anastasiou:2003ds,Melnikov:2006di,Melnikov:2006kv,Catani:2009sm,Catani:2010en}.  
One of us previously 
released a public simulation code FEWZ ({\bf F}ully {\bf E}xclusive ${\bf W}$ and ${\bf Z}$ Production) that implemented the NNLO predictions and allowed for 
arbitrary kinematic cuts to be imposed.  However, this previous version suffered from several shortcomings.  It was not easily customizable; only one cross 
section of interest could be calculated at a time, thereby necessitating numerous runs of FEWZ to obtain a kinematic histogram.  For severe cuts on the leptonic phase 
space, a Monte Carlo integration error below a few percent was not achievable~\cite{Adam:2008ge,Adam:2008pc}.
  
In this manuscript we present a new version of FEWZ which addresses the issues described above.  Specifically, the new features of FEWZ are listed below.
\begin{itemize}

\item The user can define multiple, arbitrary kinematic variables to be binned automatically during a single run.  Most of the commonly desired histograms are 
included in the new distribution of FEWZ.

\item The calculation has been broken up into 230 sectors that can be run in parallel, dramatially improving the speed and final numerical integration error.  
Sub-percent integration errors are easily obtainable even in the presence of significant phase-space restrictions.

\item For all current PDF sets, errors are automatically calculated for the total cross section and all histogram bins.

\item Most parameters of interest, such as cuts and couplings, are now set in an external input file, allowing the user complete control over the settings 
of a run.

\end{itemize}
We focus on the production of $l^+ l^-$ through $\gamma^{*}/Z$ in this manuscript; $W$ production will be addressed elsewhere.  To demonstrate the new FEWZ, 
numerous phenomenological results for the LHC are presented.  A detailed study of PDF and scale uncertainties is performed for numerous observables.  We discuss 
the effects of acceptance cuts and theoretical errors on the measurement of angular quantities in $Z$-boson events, such as the Collins-Soper angles and 
moments~\cite{Collins:1977iv,Mirkes:1992hu,Mirkes:1994dp}, and a proposed transverse-plane angular cut designed to reduce experimental backgrounds.  
A comparison with initial ATLAS and CMS results is 
performed.  We show explicitly that the technical limitations found in Ref.~\cite{Adam:2008ge} are solved.  Further details 
on how to install and run FEWZ are available in the accompanying manual.

Our manuscript is organized as follows.  In Section~\ref{sec:review}, we review the calculation of the NNLO corrections and their implementation in FEWZ.  
Section~\ref{sec:features} describes the new version of the code, with emphasis on the improvements. Phenomenological results are presented 
in Section~\ref{sec:numbers}.  We conclude in Section~\ref{sec:conc}.

\section{Review of FEWZ \label{sec:review}}

The version of FEWZ we consider here calculates the fully differential production of dilepton pairs via the neutral current (intermediate photons and $Z$-bosons).  It is designed 
to make predictions for hadron-collider observables with realistic acceptance cuts at NNLO in the strong coupling constant.  All spin correlations and finite-width effects 
are included.  The residual scale error on typical cross sections is less than 1\% and is in good agreement with the NLO scale error band.  For more 
details, we refer the reader to Refs.~\cite{Melnikov:2006di,Melnikov:2006kv}.

\subsection{Calculational details}

In QCD factorization, the Drell-Yan differential cross section for $h_1 h_2 \ra \gamma^{*},Z \ra l_1 l_2 $ can be expressed as
\be \label{diffxsect}
d\sigma = \sum_{ij} \int dx_1 dx_2 f_i^{h_1}(x_1) f_j^{h_2}(x_2) d\sigma_{ij \ra l_1 l_2}(x_1 , x_2),
\ee
where the PDFs $f_i$ represent the probability of obtaining parton $i$ from hadron $h$, and $d\sigma_{ij}$ represents the partonic cross section.  
The leading order (LO) and NLO contributions are well known~\cite{Altarelli:1979ub}.  To calculate the NNLO ($O(\alpha_S^2)$) corrections, three types of 
contributions must be considered: two-loop double-virtual contributions, one-loop real-virtual contributions with the emission of an extra parton, and 
tree-level double-real contributions, with emission of two extra partons.  Each piece is separately divergent, and must be summed to obtain a finite result.

The loop integrals of the first two types, double-virtual and real-virtual, are dealt with by decomposing the Feynman integrals into a basis of so-called master 
integrals in an automated fashion~\cite{Anastasiou:2004vj}.  These master integrals can be expressed in dimensional regularization as a Laurent series in the 
dimensional regularization parameter $\epsilon$ in terms of known functions.  The double-real contributions are also divergent, due to soft and collinear 
singularities.  The singularities are extracted by mapping each propagator denominator into a set of hypercube variables, each of which only vanishes at 
one endpoint.  Since each denominator typically vanishes when multiple variables reach an endpoint, the technique of sector decomposition~\cite{Binoth:2000ps} 
must be used to obtain the required form.  This method involves splitting the integrand into multiple terms, called {\it sectors}, which correspond to the different singular 
limits of the process. For more details for the sector decomposition required in this process, see Ref.~\cite{Anastasiou:2003gr,
Anastasiou:2005qj}.  The result of this process is a separation of the singular limits which allows them to be independently extracted.  One can perform the expansion
\be
x^{-1+\epsilon} = \frac{\delta(x)}{\epsilon} + \sum_{n=0}^{\infty} \frac{\epsilon^n}{n!}\left[\frac{\log(x)}{x}\right]_+,
\ee
where $x$ represents a hypercube variable, for each propagator topology.  The end result is approximately 200 sectors corresponding to different 
initial partons, real/virtual pieces, soft/collinear counterterms, and mappings of the hypercube variables corresponding to the singularity structure of the matrix elements.  The coefficients of the pole terms in $\epsilon$ are numerically checked to cancel when summed, and the code calculates only the $\epsilon^0$ coefficient for each sector.  The original 
calculation is more fully described in \cite{Melnikov:2006di,Melnikov:2006kv}.  We note that FEWZ incorporates only QCD corrections.  Effects such as 
final-state photonic radiation which can be important for some experimental cuts are not simulated.

\subsection{Details of the numerical code}

The resulting finite contributions are implemented in a Fortran code and interfaced with routines for calling PDF sets to complete the hadronic calculation.  
The list of supported PDF sets has been greatly expanded and will be discussed in the next section.  This results in a (4, 7, 11) dimensional integrand 
for the (LO, NLO, NNLO) computation, corresponding to the allowed degrees of freedom of the observable leptons and jets; there is an extra parameter in the 
NNLO integrand for internal purposes.  This is then interfaced with an adaptive Monte Carlo numerical integrator.  Because the kinematics of all final-state 
particles are 
reconstructed in the integrand, cuts can be imposed by checking whether requirements are met and zeroing the integrand appropriately. In the previous 
version, these cuts were input into a Fortran source file, requiring recompilation if one wants to run with a different set of cuts.  Several other parameters of interest, such as 
masses, couplings, renormalization/factorization scale, collision energy, and collision type (proton-proton vs. proton-antiproton), are set in an external 
input file in both the old and new versions of FEWZ.

For the neutral current, one is often interested in the $Z$-resonance.  In this case, a variable transformation
\be \label{Ztrans}
\frac{dM^2}{(M^2 - M_Z^2)^2 + \Gamma_Z^2 M_Z^2} = \frac{dx}{\Gamma_Z M_Z}
\ee
is made, which has the effect of flattening the $Z$-resonance, greatly improving the efficiency of numerical integration.  The new version of FEWZ allows this 
variable change to be turned on or off, depending on the region of lepton invariant mass under consideration.   To reach a numerical precision better than the errors due to scale variation (1\%), the previous 
version of the code typically required days to run for non-trivial cuts on the leptonic phase space.

\section{Description of code improvements \label{sec:features}}

We have rewritten and significantly improved FEWZ.  The new version features an improved numerical integration routine and offers users significantly more options 
for physics studies.  The resulting program is a powerful tool that enables precision studies of most aspects of lepton-pair production at hadron colliders.  
We describe in this section the new features of FEWZ.  More details on running FEWZ can be found in the accompanying manual.

\subsection{Overview of improvements}

Our neutral-current code has been updated to improve speed and precision, as well as increase the amount of information generated during a single run.  We provide a summary of the changes and their intended effect in this section.  While we focus here on neutral-current production, the same improvements apply also to $W$ production.  This will be addressed in a future publication.
\begin{description}

\item {\bf Parallelization:} Each of the 230 NNLO sectors is calculated independently.  This allows the Monte Carlo integration to adapt to the structure 
of a single sector rather than to all at once, and also allows more than one processor core to work simultaneously.
We have written scripts for starting multi-core local runs, and also for running on the \emph{Condor}~\cite{condor} distributed computing system, as well as combining results from individual sectors.

\item {\bf Run parameters:} All inputs, including cuts on leptons and jets, electroweak couplings, and other parameters which control run setting, are now set in an external input file, allowing the user complete flexibility to customize FEWZ.

\item {\bf Histograms:} By tabulating the weights associated with each event, kinematic distributions are now produced automatically during a run, with little overhead.  The user can select which histograms to fill in an external input file.  Most distributions of interest are included in the default version of FEWZ.

\item {\bf PDF errors:} When running with PDF sets that contain error eigenvectors, all eigenvectors are calculated automatically for each histogram bin.  The resulting output can be combined using the included scripts to produce a final output file that contains the integration error as well as PDF error for both the total cross section and each histogram bin.

\end{description}

\subsection{Details of the numerical integration}

Quantities of interest, such as the total cross section and its various kinematical distributions, are produced by numerically integrating Eq.~(\ref{diffxsect}) with a Monte Carlo adaptive integrator.  We use the standard \emph{Vegas} routine from the package Cuba 1.7~\cite{Hahn:2004fe}, which is distributed with our program.  \emph{Vegas} allows one to save the state of the integration between grid adaptations, which is useful for long calculations and allows a pause so that intermediate output is produced.  In addition, the weight of each sampling point is returned, which we use to calculate histograms bins as described later.

Several additional variable transformations are implemented to improve the performance of the numerical integration.  In addition to the removal of the Z propagator using Eq.~(\ref{Ztrans}), an additional smoothing of certain sectors' integrands is performed.  After the $\epsilon$-singularities are removed, the integrands may still diverge logarithmically at 0 or 1 in the hypercube variables, even though the integrals are still finite.  For stability of the adaptive integrator, transformations such as
\begin{equation}
dx \rightarrow 6 u (1 - u) du
\end{equation}
are performed for the NNLO sectors.  Such a transformation removes singularities of the form ${\rm ln}(x)$ while restricting 
the integration region's support to the unit hypercube.

\subsection{Parallelization}

As mentioned, the integrand has been broken into 230 sectors corresponding roughly to the results of sector decomposition.  Some sectors, however, have been merged or split.  After sector decomposition, there were approximately 260 sectors; these correspond to different initial state partons (gluon-gluon, gluon-quark, quark-quark), different diagram type (real-real, real-virtual, soft/collinear counterterms), and different decomposition of the singularity structure.  A dummy run of these sectors was set up for a typical input.  It was found that some sectors anti-correlate over the randomly generated phase space points.  For the results of such sectors such $i$ and $j$, this implies
\begin{equation}
Cov(i, j) < \sigma_i \sigma_j ,
\end{equation}
where $\sigma_i$ denotes the standard deviation of the random sample of integrand $i$, and $Cov(i,j)$ the covariance of the two sectors.  This indicates that these sectors should be combined, since the resulting error would be smaller than random sampling each independently, {\it i.e.}, they are canceling.  Candidate combinations were tested with multiple sets of cuts, and those that performed better together were combined.  Keeping the remainder separate allows \emph{Vegas} to adapt its grid to the different integrand shapes better and accelerates convergence.

A few sectors typically take much longer to reach a target error than others due to high variance and evaluation time. To prevent these from holding up the user in a cluster environment where numerous processors are available, identical copies of these extreme sectors are split over multiple new sectors with differing random seeds to statistically reduce their contribution to the total error.  Each sector is given a target precision to reach depending on the total absolute precision set in the input file.  In the worst-case where the integrator fails to adapt ($\epsilon_i \simeq \sqrt{V_i/N}$), it is found to be most CPU-efficient to weight the goal error for each sector according to
\begin{equation}
\epsilon_i^2 = \frac{\sqrt{V_i t_i}}{\sum_j \sqrt{V_j t_j}} \epsilon_{tot}^2
\end{equation}
where $V_i$ is the sector's random-sampling variance (estimated for default cuts) and $t_i$ is the evaluation time per point.

The sector number can be set in the input file and run with the compiled program, which will then produce a human-readable output file.  However, it is intended that the user instead run a provided starting script which will create a directory structure for the results of all sectors and run the code on all sectors.  There is a script for running locally on a specific number of cores, and one for running on the \emph{Condor} system, where a job is submitted for each sector.  The user can then run a combining script once output files have been produced for each sector; this only requires that one iteration of \emph{Vegas} has completed.  Further details are available in the FEWZ manual.

\subsection{Run parameters}

All relevant input parameters are now set in an external input file.  These include vector-boson masses and couplings of the photon and $Z$ to fermions.  The vector and axial couplings of the 
$Z$ are set separately from $\alpha_{QED}$ and ${\rm sin}^2 \theta_W$, allowing an arbitrary electroweak scheme or an improved Born approximation to be implemented.  The user may also select whether to optimize the integration for the $Z$-peak as described in Eq.~(\ref{Ztrans}).  The new version of FEWZ allows numerous cuts to be selected in the input file.  These include restrictions on the following quantities:
\begin{itemize}
\item lepton transverse momenta and the dilepton transverse momentum;
\item lepton pseudorapidities and the dilepton rapidity;
\item dilepton invariant mass;
\item jet transverse momenta and pseudorapidities;
\item the number of observable jets;
\item jet-jet, jet-lepton, and lepton-lepton isolation.
\end{itemize}
The input file also allows one to select the jet-merging algorithm (cone or anti-$k_T$), as well as the chosen PDF set.  We detail in Sec.~\ref{subsec:PDFs} the supported PDFs.

\subsection{Histogramming}

Since each point in the 11-dimensional parameter space generated by \emph{Vegas} corresponds to particular kinematics, we can save this information to reconstruct more detailed distributions than just the total cross section, with little overhead.  The ``event" (whose contribution may be negative) is sorted into the appropriate bin for each of the many distributions defined.  The bin size and histogram extent can be changed by the user in a histogram input file.  Since \emph{Vegas} returns the weight of the generated point, this weight is used to keep track of the weighted average and standard deviation for each bin in the same fashion as \emph{Vegas} for the total.  A copy of each histogram bin is made for each PDF error eigenvector, reweighted appropriately, for determining PDF errors later.  All of this information is stored in a file updated after each \emph{Vegas} iteration, in case a calculation needs to be stopped or restarted.

A large number of histograms have been included in the default version of FEWZ.  Some differential distributions provided include:
\begin{itemize}
\item dilepton transverse momentum;
\item dilepton rapidity;
\item dilepton invariant mass;
\item lepton transverse momenta;
\item lepton pseudorapidities;
\item jet transverse momenta;
\item jet pseudorapidites;
\item $\Delta R$ separation between observable particles;
\item $H_T$ (scalar sum of all transverse momenta);
\item $\cos(\theta^{*})$, the lepton polar angle in the Collins-Soper frame~\cite{Collins:1977iv};
\item $\Delta \phi$, the lab-frame transverse-plane angular separation between leptons.
\end{itemize}
In addition, by reweighting each phase-space point according to certain trigonometric functions in the Collins-Soper coordinates, we can reconstruct the Collins-Soper moments $A_i$~\cite{Collins:1977iv,Mirkes:1992hu,Mirkes:1994dp}, binned in dilepton transverse momentum.  We currently support only one-dimensional histograms in FEWZ.  Since the weights 
of all events are saved, it is straightforward to extend the program to handle higher-dimensional histograms. 

\subsection{PDFs} \label{subsec:PDFs}

The number of supported PDF sets has been drastically expanded from the previous version of FEWZ.  All modern distributions are now available.  At present, the code supports and includes the following sets:
\begin{itemize}
\item ABKM 09 NLO and NNLO~\cite{Alekhin:2009vn};
\item CTEQ versions 6L1~\cite{Pumplin:2002vw}, 6.5~\cite{Tung:2006tb}, 6.6~\cite{Nadolsky:2008zw}, 10 and 10W~\cite{Lai:2010vv};
\item GJR 08 LO/NLO~\cite{Gluck:2008gs} and JR 09 NNLO~\cite{JimenezDelgado:2009tv};
\item MRST 2006 NNLO~\cite{Martin:2007bv} and MSTW 2008 LO/NLO/NNLO~\cite{Martin:2009iq};
\item NNPDF2.0~\cite{Ball:2010de}.
\end{itemize}
All sets but CTEQ 6L1 include PDF error eigenvectors.  The results for each eigenvector set are calculated and stored in parallel to the central PDF by reweighting each generated phase space point appropriately.  The results are stored in an auxiliary output file.  When the results of all sectors are combined, the PDF errors are calculated by summing the eigenvectors in quadrature (symmetric sets), finding the standard deviation from the mean (neural network sets), or with the procedure described in Ref.~\cite{Tung:2006tb} (asymmetric sets).

\subsection{Additional Features}

We have provided shell scripts for farming out the sectors in parallel either locally or on \emph{Condor}, and a finishing script which combines the results of individual sectors.  In addition to the basic operation of combining the sectors and computing PDF errors, the finishing script can perform operations such as addition, subtraction, multiplication, and division on different runs, all while treating the integration and PDF errors consistently.  This can be useful for computing the results of disjoint cuts, calculating K-factors, acceptances, and PDF correlations between different kinematic regions.  These features are described in more 
detail in the manual distributed with FEWZ.

\subsection{Runtime benchmarks \label{sec:benchmarks}}

To demonstrate the performance of the current version of our code, we compare it against the previous version by running to various target precisions under typical usage.  For the following test we use the standard set of cuts described in Sec.~\ref{sec:numbers}, minus the isolation cuts that were unavailable in the previous version.  MSTW 2008 NNLO PDF sets are used, but the results are typical for all PDF sets, as PDF evaluation calls require an insignificant amount of CPU time.  The result is a cross section of approximately 440 pb, leading to an acceptance of 46\%.  These benchmarks are run on an 8-core (2 CPU) Intel Xeon E5335 2.0 GHz machine running Scientific Linux 5.5; the new version of the code is able to use the multiple cores available on modern hardware.  In addition to calculating the cross section with cuts, the new version is simultaneously doing all of the histogramming and PDF error eigenset running described above; all histograms available have been activated.  The integration error versus time is shown for both versions below in Fig.~\ref{errorbench}.

\begin{figure}[h!]
\begin{center}
\includegraphics[width=4.0in]{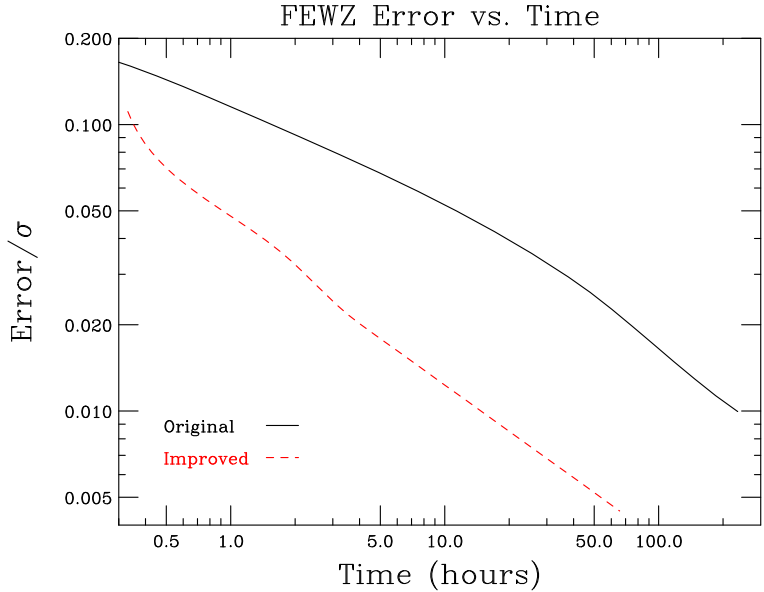}\\
\caption{Relative error versus time for the previous version of FEWZ and for the current version.
\label{errorbench}}
\end{center}
\end{figure}

For a target precision of 1\% and with standard cuts, the previous version of the code required 230 hours (9.8 days) and reached a 0.97\% relative error.  The  new version required 20 hours and reached a 0.84\% error.  The new version outperforms the old, even on a per-core basis, all while computing over a dozen kinematic distributions and the PDF errors for each.  The detailed results for phenomenological quantities are presented in the next section.

\section{Phenomenological results \label{sec:numbers}}

We present in this section phenomenological results for the LHC that illustrate the improvements in FEWZ detailed in previous sections.  Many predictions shown have 
not previously been presented in the literature.  All numerical results shown use the $G_{\mu}$ electroweak scheme~\cite{Dittmaier:2001ay}. The numerical values for 
the various parameters are shown below:
\begin{eqnarray}
M_Z &=& 91.1876 \,\, {\rm GeV},\;\;\; M_W = 80.403 \,\, {\rm GeV}, \nonumber \\
\Gamma_Z &=& 2.4952 \,\, {\rm GeV},\;\;\; \Gamma_{Z \to l^+ l^-} = 0.08399 \,\, {\rm GeV},\;\;\;
G_{\mu} = 1.16637 \times 10^{-5}\,\, {\rm GeV}^{-2}.
\end{eqnarray}
We use 7 TeV for the LHC energy, and use the value of $\alpha_S(M_Z)$ dictated by the appropriate PDF set.  

\subsection{Benchmark numbers}

We begin by presenting results for the {\it inclusive} cross section at NNLO, which we define with only the invariant mass cut $66 \, {\rm GeV} 
\leq M_{ll} \leq 116 \, {\rm GeV}$.  To define the error arising from scale variation, we let the factorization and renormalization scales vary 
separately in the range 
$M_Z /2 \leq \mu_{R,F} \leq 2 M_Z$ subject to the restriction $1/2 < \mu_R / \mu_F < 2$.  The PDF error is defined using the procedure recommended by the 
group which produced the fit.  We only use the most recent PDF sets for which an NNLO extraction is available: MSTW 2008, ABKM 
2009, and JR 2009.  We find the following 
results for the MSTW set, with all sources of error indicated:
\begin{equation}
\text{MSTW 2008:} \;\; \sigma_{inc}=963.7^{+4.9}_{-6.8} \text{(scale)} ^{+33.7}_{-30.0}\text{(PDF)}\pm 0.5 \text{(tech.)} \; \text{pb}.
\end{equation}
We have included as the final error component the technical precision that arises from the Vegas integration.  This number is below $ \pm 0.1\%$, and does not 
significantly affect either the central value or the estimate of the error.  The PDF uncertainty is the dominant component 
of the error.  We now compare this with the central values and errors obtained using the other NNLO PDF sets:
\begin{eqnarray}
\text{MSTW 2008:}&& \sigma_{inc}=963.7^{+33.7}_{-30.0}\text{(PDF)}\; \text{pb}; \nonumber \\
\text{ABKM 2009:}&& \sigma_{inc}=980.5^{+15.6}_{-15.6}\text{(PDF)}\; \text{pb}; \nonumber \\
\text{JR 2009:}&& \sigma_{inc}=907.3^{+17.9}_{-20.2}\text{(PDF)}\; \text{pb}.
\end{eqnarray}
The relative scale and technical errors are similar to the MSTW values, and are negligible compared to the PDF uncertainties.  We note that the 90\% C.L. error 
bands are shown for the MSTW fit, while the ABKM and JR quoted errors should be interpreted as $1\sigma$ uncertainties.  Upon scaling the MSTW uncertainties 
down by a factor of 1.6, they agree approximately with those of the ABKM and JR fits.  MSTW and ABKM give similar predictions 
for the central value.  The JR 2009 set gives a central value 6\% lower than MSTW 2008 and 
8\% lower than ABKM 2009.  These results for the inclusive cross section can be compared to preliminary results from ATLAS~\cite{atlasresult} and CMS~\cite{cmsresult}:
\begin{eqnarray}
\text{ATLAS:} && \sigma_{inc} = 830^{+70}_{-70} \text{(stat.)} ^{+60}_{-60} \text{(syst.)}  \pm 90 \text{(lumi.)} \; \text{pb}; \nonumber \\
\text{CMS:} && \sigma_{inc} = 882^{+77}_{-73} \text{(stat.)} ^{+42}_{-36} \text{(syst.)}  \pm 97 \text{(lumi.)} \; \text{pb}.
\end{eqnarray}
The ATLAS result agrees with the theoretical prediction for all three PDF sets within the current errors.  
We note that the CMS cross section is defined with the invariant mass cut $60 \, {\rm GeV} \leq M_{ll} \leq 120 \, {\rm GeV}$, while ATLAS uses the same cut as our 
default choice.  Changing the mass window to match the CMS definition leads to a 1.5\% increase in our result for all three PDF sets; for example, 
our MSTW cross section becomes
\begin{equation}
\text{MSTW 2008:}\;\; \sigma_{inc}=977.7^{+34.2}_{-30.5}\text{(PDF)}\; \text{pb},
\end{equation}
which agrees with the CMS measurement within errors.

We now define the {\it standard acceptance cuts}, which include the following restrictions in addition to the previous invariant mass cut:
\begin{eqnarray}
p_{T,lep} &>& 25 \,\, \text{GeV},\;\;\; |\eta_{lep}| < 2.5, \nonumber \\
\Delta R_{lep,lep} &>& 0.5,\;\;\; \Delta R_{lep,jet}>0.5.
\label{standardacc}
\end{eqnarray}
We have used the standard definition $\Delta R_{12} = \sqrt{(\eta_1-\eta_2)^2+(\phi_1 - \phi_2)^2}$ in defining the isolation cuts on the leptons and jets.  The partons have 
been clustered according to the anti-$k_t$ algorithm with separation parameter $R=0.5$.  We note that the simple $\Delta R_{lep,jet}$ cut imposed here is not how the isolation 
requirement is implemented experimentally.  This effect would also typically be assigned to the efficiency rather than the acceptance.  We include the cut here to demonstrate the 
ability of FEWZ to reconstruct jets and constrain hadronic activity.  The effect of the isolation requirement is to reduce the cross section by only a couple of percent, and this cut can 
be easily removed if desired.  The cross sections after these cuts have been imposed are as follows:
\begin{eqnarray}
\text{MSTW 2008:}&& \sigma_{standard}=436.0^{+15.4}_{-13.9}\text{(PDF)} \; \text{pb}; \nonumber \\
\text{ABKM 2009:}&& \sigma_{standard}=445.6^{+7.6}_{-7.6}\text{(PDF)}\; \text{pb}; \nonumber \\
\text{JR 2009:}&& \sigma_{standard}=404.3^{+7.9}_{-11.0}\text{(PDF)}\; \text{pb}.
\end{eqnarray}
The relative scale and technical errors are similar to the inclusive case and are much smaller than the PDF errors, and have not been included.  By taking the ratio of this over the inclusive rate, the 
acceptance can be derived.  The scripts provided with FEWZ allow this to be done for each error eigenvector to obtain the PDF error.  We find the following for each 
set:
\begin{eqnarray}
\text{MSTW 2008:}&& A_{standard}=0.4525^{+0.0033}_{-0.0040}\text{(PDF)}; \nonumber \\
\text{ABKM 2009:}&& A_{standard}=0.4544^{+0.0018}_{-0.0018}\text{(PDF)}; \nonumber \\
\text{JR 2009:}&& A_{standard}=0.4456^{+0.0027}_{-0.0039}\text{(PDF)}.
\end{eqnarray}
The technical precisions for each acceptance are well below $\pm 0.1\%$.  The PDF errors on the acceptance are much smaller than for the individual cross sections, as 
expected, and range from 0.4\% for ABKM 2009 to approximately 0.8\% for MSTW 2008.  The results for ABKM and MSTW are in good agreement, but the JR 2009 acceptance is 
1.5\% lower than that for MSTW, larger than the estimated 90\% C.L. MSTW PDF error by a factor of two.  Scale errors for such cuts 
have previously been studied in Ref.~\cite{Melnikov:2006kv}, and are at the percent-level or below for both the cross section and acceptance.

To address the limitations imposed by significant phase-space restrictions, we define a {\it severe acceptance cut} following the analysis in Ref.~\cite{Adam:2008ge}:
\begin{eqnarray}
p_{T,lep} &>& 25 \,\, \text{GeV},\;\;\; 1.5 < |\eta_{lep}| < 2.3, \nonumber \\
\Delta R_{lep,lep} &>& 0.5,\;\;\; \Delta R_{lep,jet}>0.5.
\end{eqnarray}
Only a small slice of the forward region in lepton pseudorapidity is taken.  In the study of Ref.~\cite{Adam:2008ge} using the old version of FEWZ, an integration 
precision of only $\pm 3\%$ was obtainable after asymptotic running, preventing an accurate estimate of the higher-order 
corrections in this phase-space region.  
Using the new version of FEWZ, we obtain a technical precision at the $\pm 0.5\%$ level after several days of running:
\begin{equation}
\text{MSTW 2008:} \;\; \sigma_{severe}=37.09^{+0.54}_{-0.86} \text{(scale)} ^{+1.24}_{-1.22}\text{(PDF)}\pm 0.18 \text{(tech.)} \; \text{pb}.
\end{equation}
We note that the difference between the central value presented here and in Ref.~\cite{Adam:2008ge} is due primarily to the 7 TeV energy we use.  The technical precision is 
a factor of a few less than the (small) scale dependence, even though the acceptance is only 4\% for this cut.  We view this as evidence that the limitations 
imposed by integration errors are solved for all studies of interest at the LHC.  The PDF errors are again the dominant uncertainty on this result, leading us 
to study the results for the other NNLO PDF sets:
\begin{eqnarray}
\text{MSTW 2008:} && \sigma_{severe}= 37.09^{+1.24}_{-1.22}\text{(PDF)}; \nonumber \\
\text{ABKM 2009:} && \sigma_{severe}=36.85^{+0.59}_{-0.59}\text{(PDF)}; \nonumber \\
\text{JR 2009:} && \sigma_{severe}=35.20^{+0.78}_{-0.81}\text{PDF}.
\end{eqnarray}
These lead to the following results for the acceptances and associated PDF errors:
\begin{eqnarray}
\text{MSTW 2008:}&& A_{severe}=0.03835^{+0.00029}_{-0.00033}\text{(PDF)}; \nonumber \\
\text{ABKM 2009:}&& A_{severe}=0.03752^{+0.00015}_{-0.00015}\text{(PDF)}; \nonumber \\
\text{JR 2009:}&& A_{severe}=0.03880^{+0.00039}_{-0.00043}\text{(PDF)}.
\end{eqnarray}
The conclusion of these studies is that FEWZ is capable of providing results with sub-0.5\% integration errors even in the presence of severe phase-space 
restrictions, allowing central values, scale and PDF errors to be accurately computed for observables of interest at the LHC.

\subsection{Results for distributions: {\it inclusive} cuts}

\begin{figure}[h!]
\begin{center}
\includegraphics[width=4.75in]{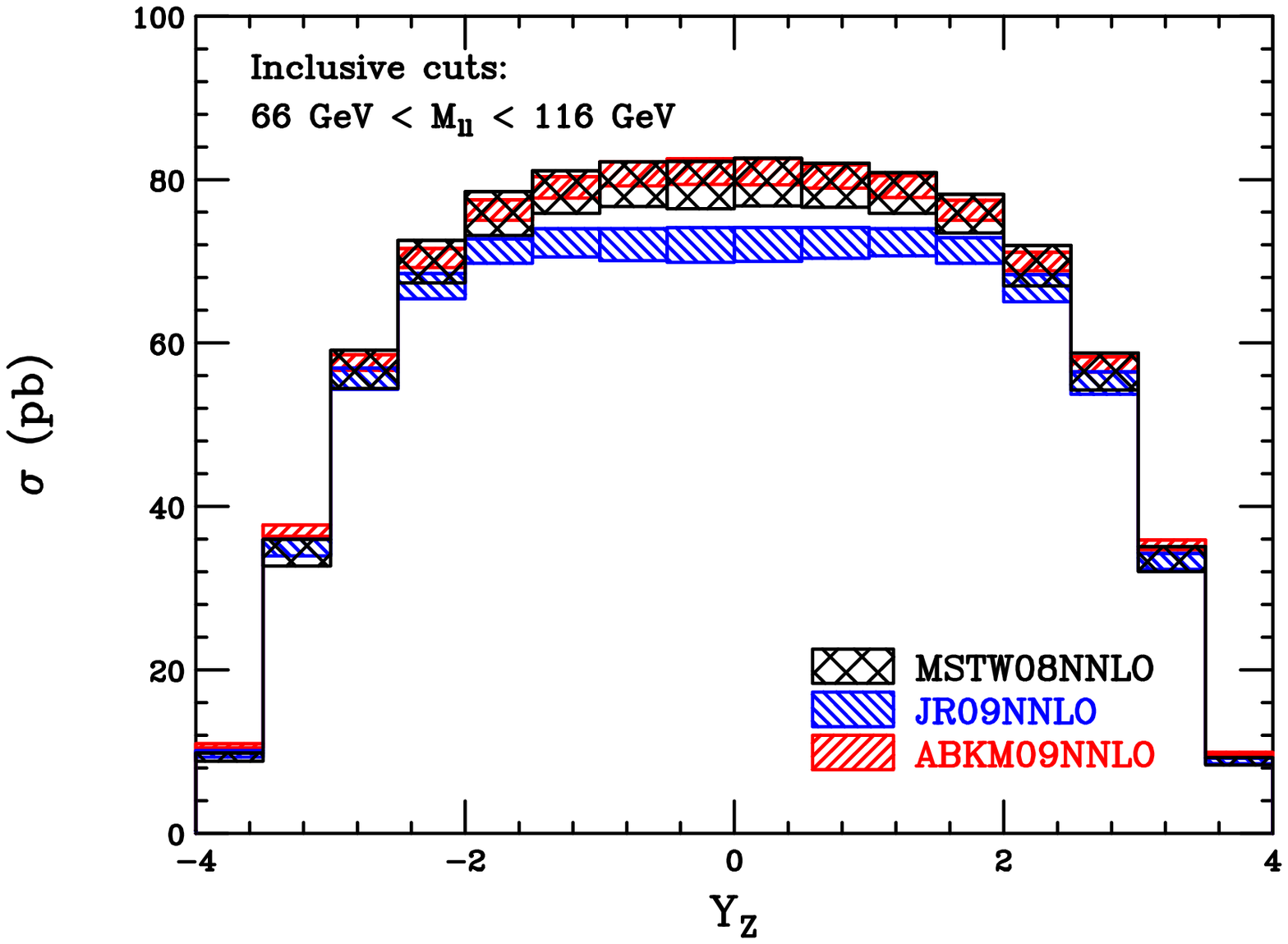}\\
\vspace{0.35cm}
\includegraphics[width=4.75in]{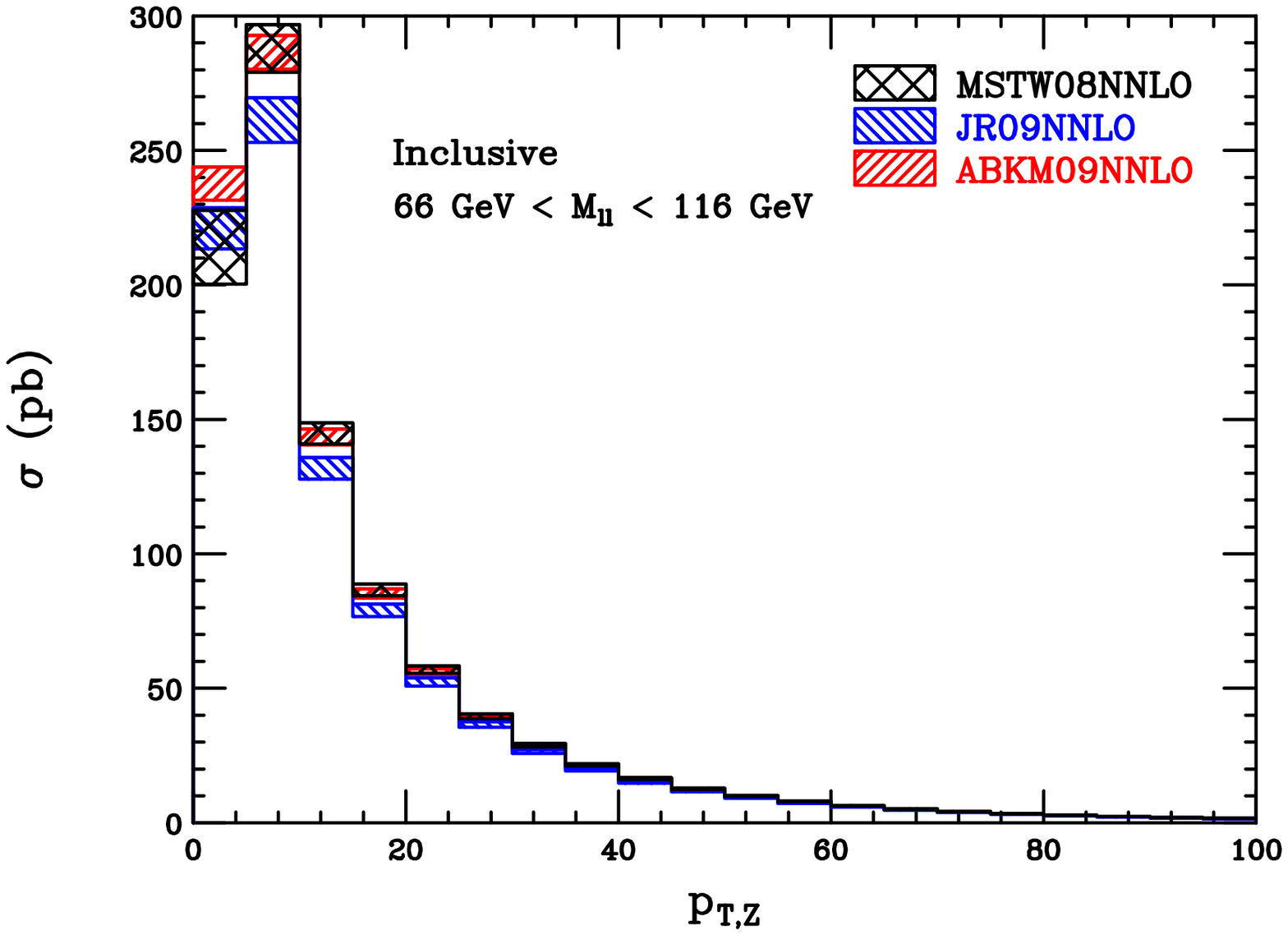}
\caption{Bin-integrated cross sections for the lepton-pair rapidity (upper panel) and transverse momentum (lower panel) for all three NNLO PDF sets.  
Only a cut on the invariant mass $66 \, {\rm GeV} \leq M_{ll} \leq 116 \, {\rm GeV}$  has been implemented.  The bands indicate the PDF uncertainties for each set.
\label{Zbasic}}
\end{center}
\end{figure}

\begin{figure}[h!]
\begin{center}
\includegraphics[width=4.75in]{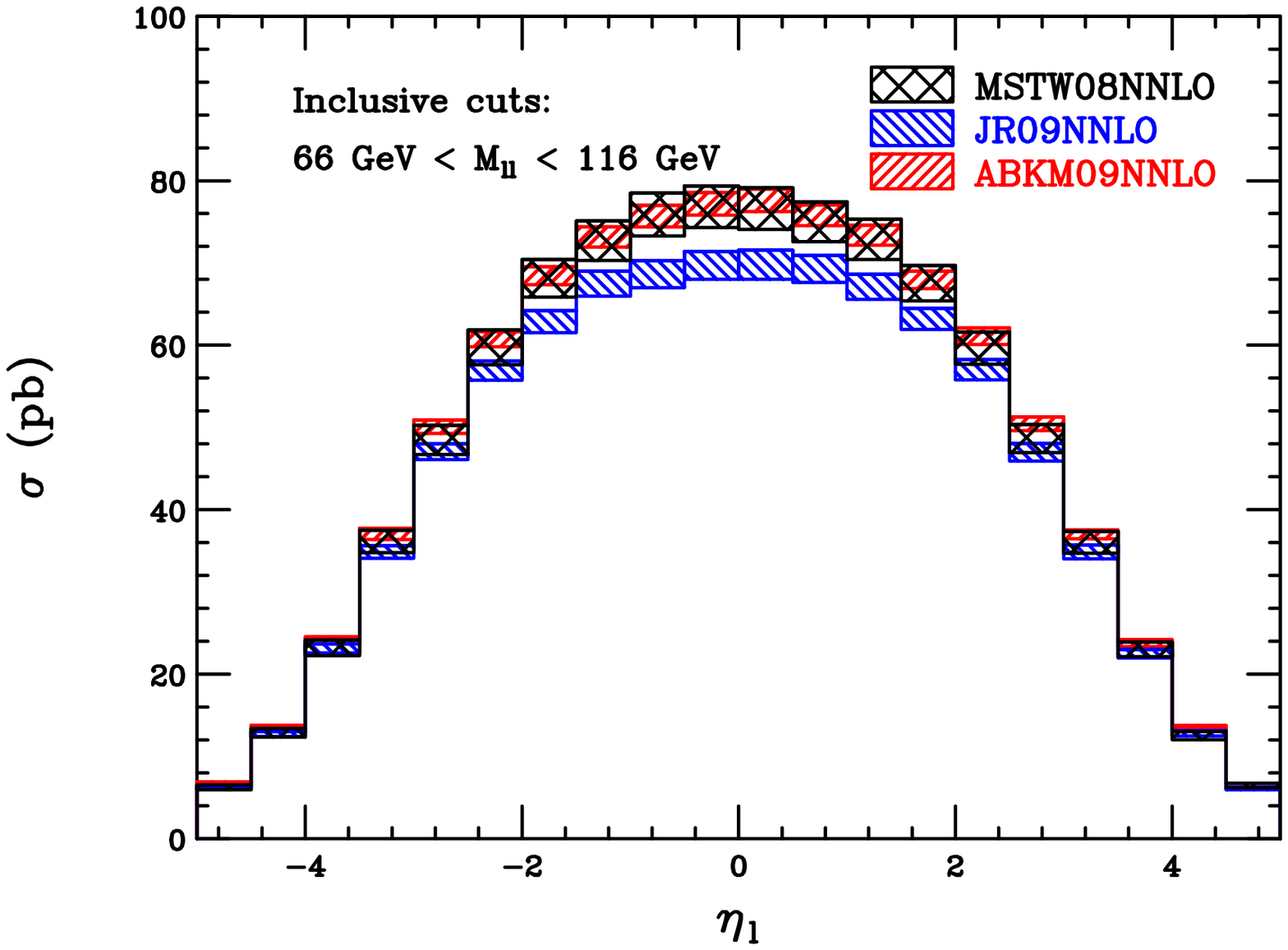}\\
\vspace{0.35cm}
\includegraphics[width=4.75in]{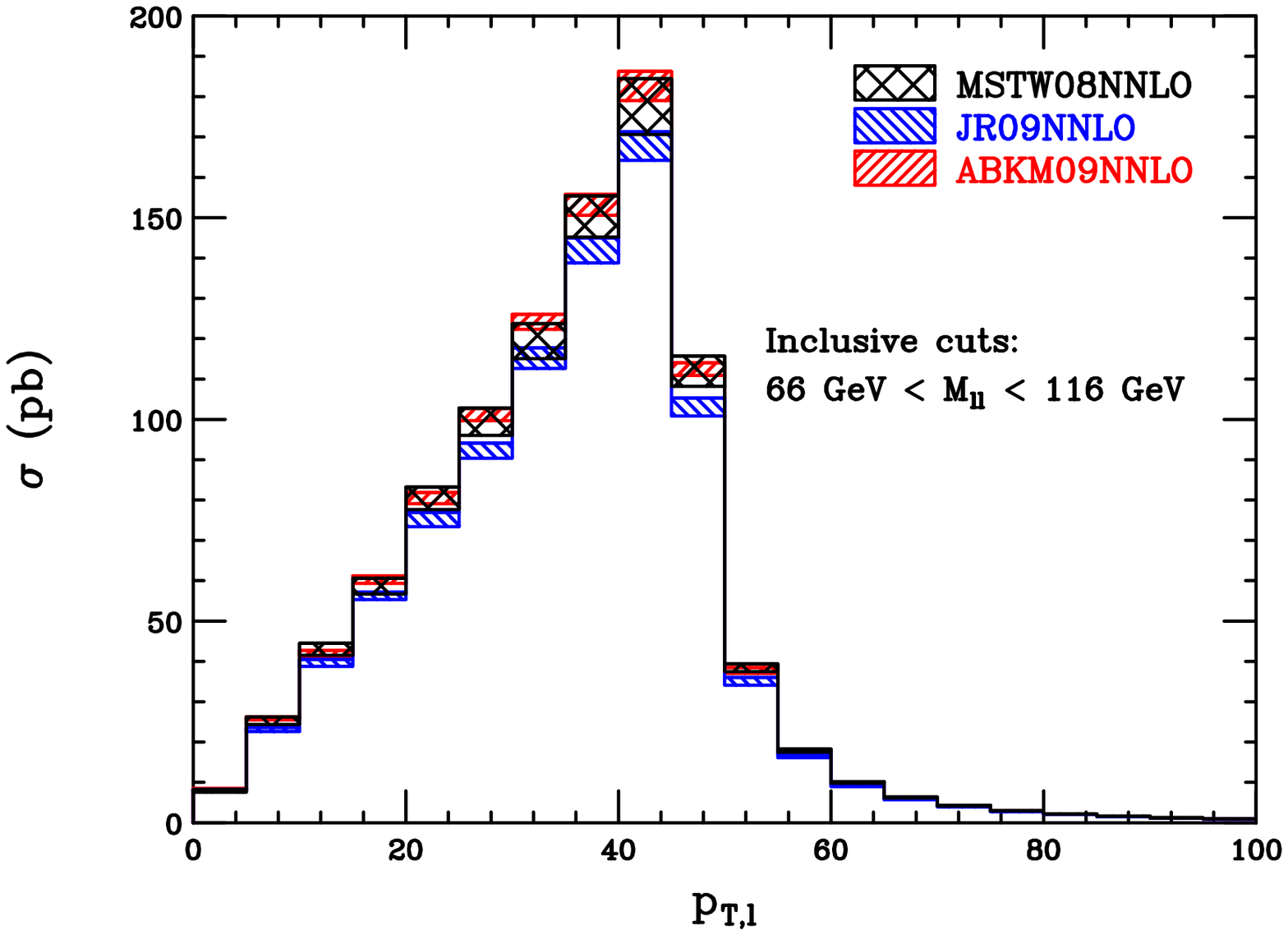}
\caption{Bin-integrated cross sections for the lepton pseudorapidity (upper panel) and transverse momentum (lower panel) for all three NNLO PDF sets.  
Only a cut on the invariant mass $66 \, {\rm GeV} \leq M_{ll} \leq 116 \, {\rm GeV}$  has been implemented.  The bands indicate the PDF uncertainties for each set.
\label{lepbasic}}
\end{center}
\end{figure}

We now present results for LHC distributions, both to demonstrate the histogramming features of the code and also to present several new phenomenological results.  
We begin by running FEWZ once for each NNLO PDF set in the {\it inclusive} mode, with only a cut on the invariant mass $ 66 \, {\rm GeV} 
\leq M_{ll} \leq 116 \, {\rm GeV}$ imposed.  Bin-integrated cross sections for several standard kinematic distributions of the lepton pair are 
shown in Fig.~\ref{Zbasic}.  Integration errors reach a maximum of 1\% for each bin, and are much below this 
value except near phase-space edges.  
Scale errors are small, and the dominant uncertainties come from PDFs.  Only these are shown in the figure, and are indicated by the hatched bands.  The smaller 
integrated result for the 
JR 2009 distribution is apparent from both plots.  The rapidity distribution of the reconstructed $Z$ is also slightly flatter for the JR 2009 set.  The 
ratio between the first and second bins of the $p_T$ distribution differs for each set by an amount larger than the estimated uncertainty.  However, this 
bin only includes the range $0 \, {\rm GeV} \leq p_{T,Z} \leq 5 \, {\rm GeV}$, and fixed-order perturbation theory is not expected to accurately describe this 
region.  If this ratio difference persists after the including the resummation of low-$p_{T,Z}$ logarithms, it could be an interesting discriminator between 
different PDF extractions.  In addition to distributions of dilepton variables, FEWZ also produces histograms of leptonic variables.  Results for several 
basic leptonic distributions are shown in Fig.~\ref{lepbasic}.  The Jacobian peak present at $p_{T,l}=45$ GeV is visible in the transverse momentum spectrum, 
as is the tail generated by additional QCD radiation beginning at ${\cal O}(\alpha_s)$.  Since the lepton-pair transverse momentum distribution and the 
lepton transverse momentum distribution above the Jacobian peak are generated by additional QCD radiation, their perturbative expansion begins at ${\cal O}(\alpha_s)$ and FEWZ effectively produces only NLO distributions for these observables.  
The corrections for these quantities could also be obtained from an NLO calculation for $Z$+1 jet, such as implemented in MCFM~\cite{mcfm}.

\begin{figure}[h!]
\begin{center}
\vspace{0.2cm}
\includegraphics[width=4.5in]{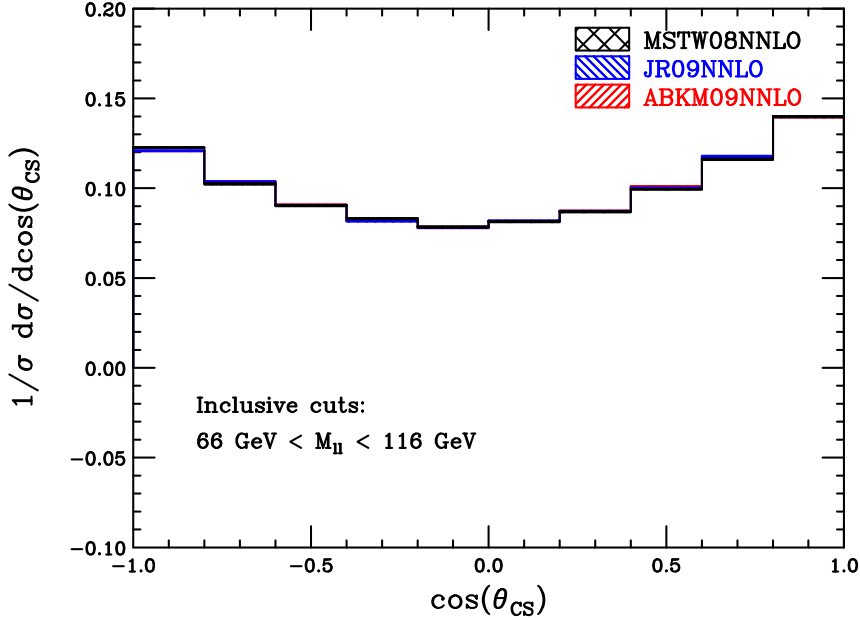}
\caption{Bin-integrated, normalized ${\rm cos}\,\theta$ distribution for all three NNLO PDF sets.  The polar angle of the lepton is defined in the 
Collins-Soper frame, as indicated by the subscript in the plot.  Only a cut on the invariant mass $66 \, {\rm GeV} \leq M_{ll} \leq 116 \, {\rm GeV}$  
has been implemented.  The bands indicate the PDF uncertainties for each set.  We note that the smallness of the PDF errors makes it difficult to distinguish 
the three separate bands in the plot.
\label{CSctheta}}
\end{center}
\end{figure}

\begin{figure}[h!]
\begin{center}
\vspace{0.2cm}
\includegraphics[width=4.5in]{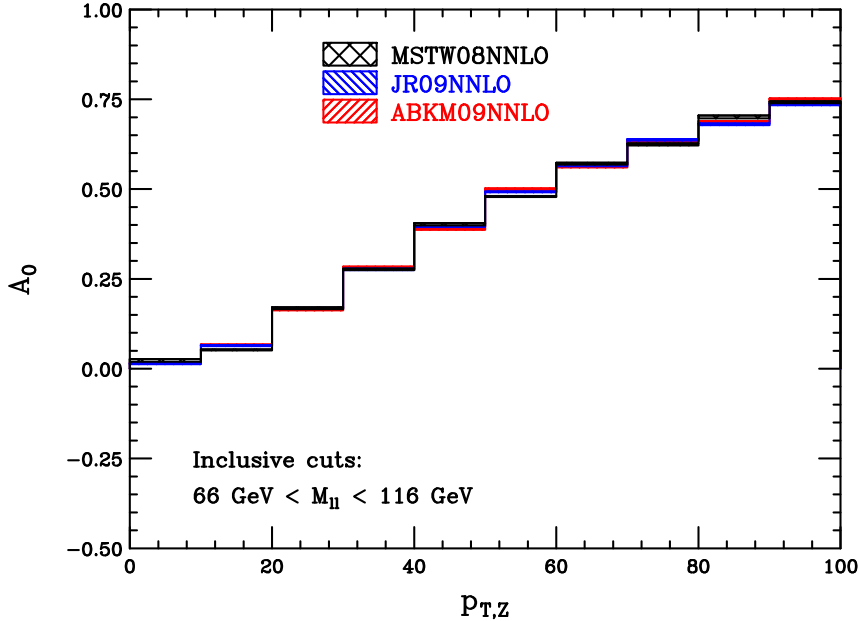}
\caption{Bin-integrated results for the CS moment $A_0$, presented as a function of the lepton-pair transverse momentum, 
for all three NNLO PDF sets.  
Only a cut on the invariant mass $66 \, {\rm GeV} \leq M_{ll} \leq 116 \, {\rm GeV}$  has been implemented.  The bands indicate the PDF uncertainties for each set. We note that the smallness of the PDF errors makes it difficult to distinguish 
the three separate bands in the plot. 
\label{CS1}}
\end{center}
\end{figure}

\begin{figure}[h!]
\begin{center}
\includegraphics[width=4.5in]{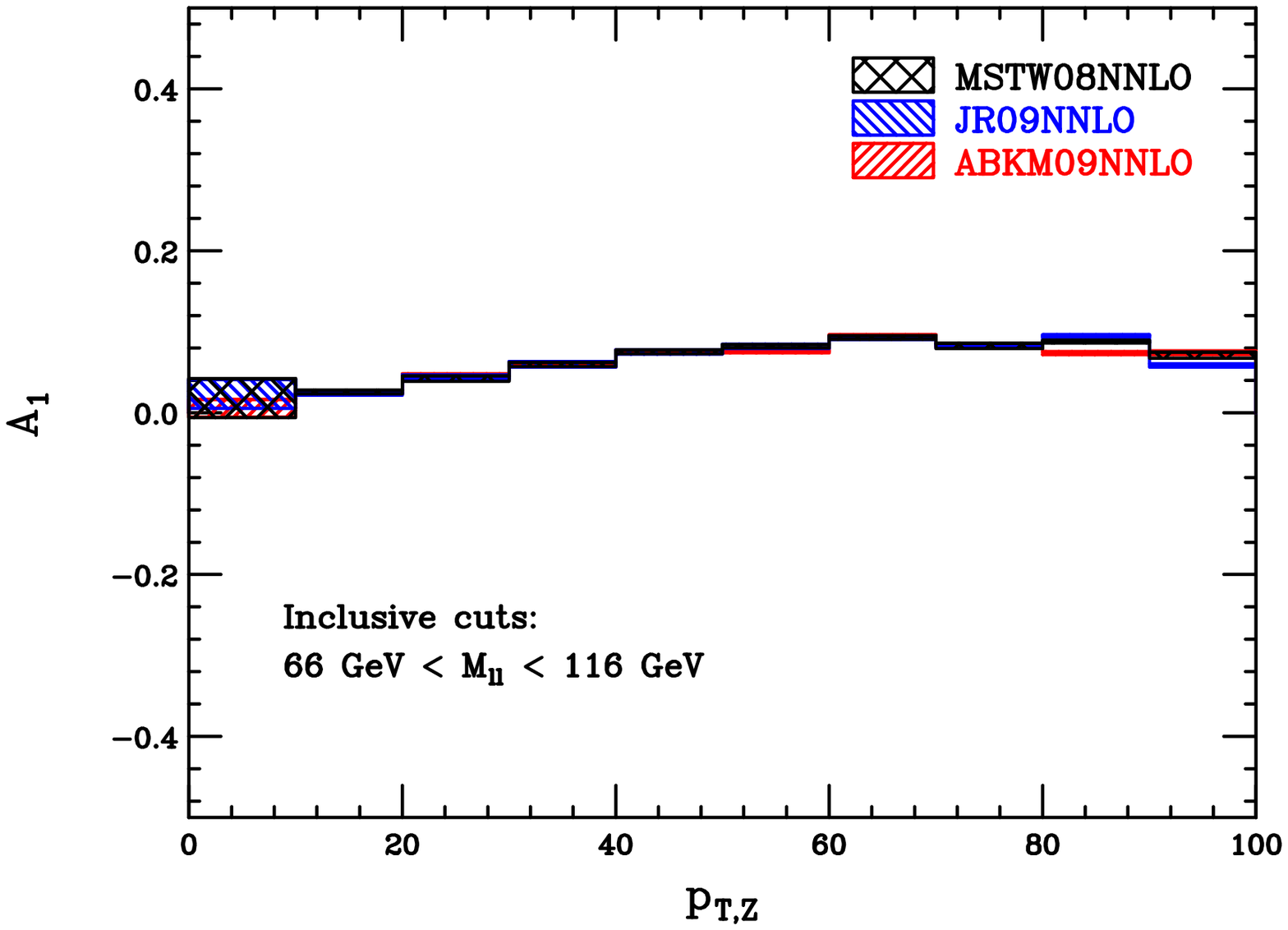}\\
\vspace{0.35cm}
\includegraphics[width=4.5in]{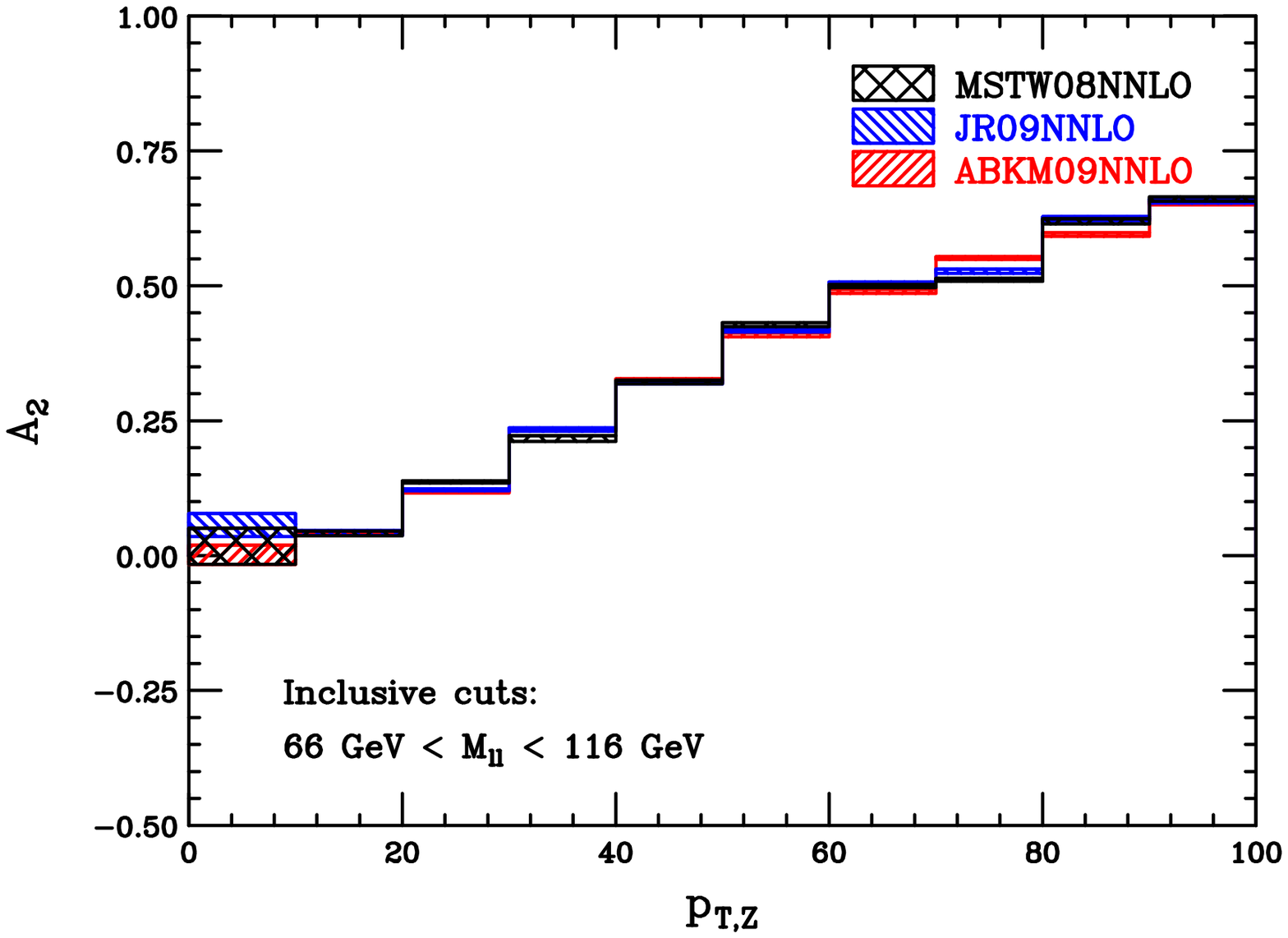}
\caption{Bin-integrated results for the CS moments $A_1$ (upper panel) and $A_2$ (lower panel), presented as a function of the lepton-pair transverse momentum, 
for all three NNLO PDF sets.  
Only a cut on the invariant mass $66 \, {\rm GeV} \leq M_{ll} \leq 116 \, {\rm GeV}$  has been implemented.  The bands indicate the PDF uncertainties for each set.  We note that the smallness of the PDF errors makes it difficult to distinguish 
the three separate bands in the plot.
\label{CS2}}
\end{center}
\end{figure}

\begin{figure}[h!]
\begin{center}
\includegraphics[width=4.6in]{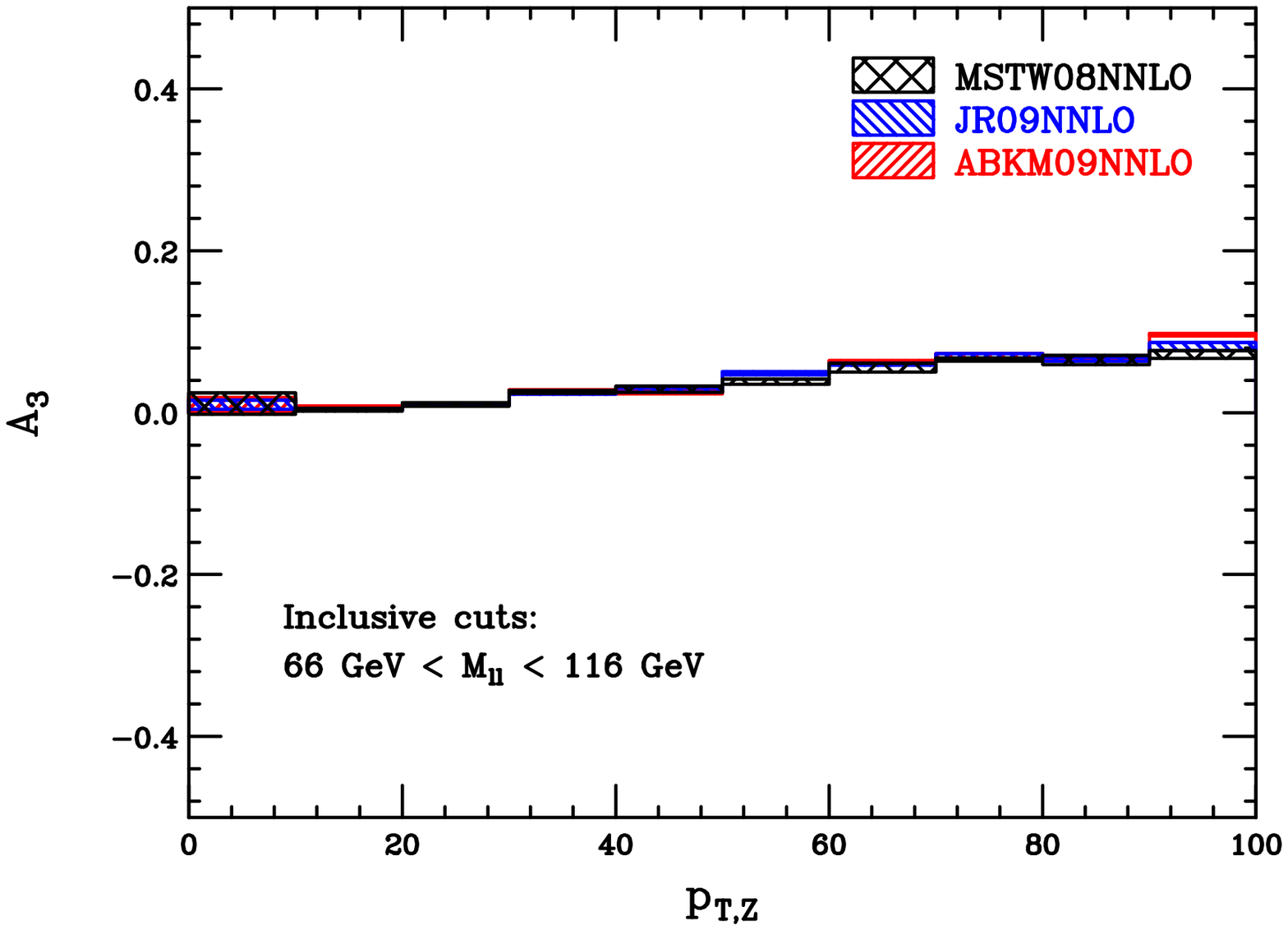}\\
\vspace{0.35cm}
\includegraphics[width=4.6in]{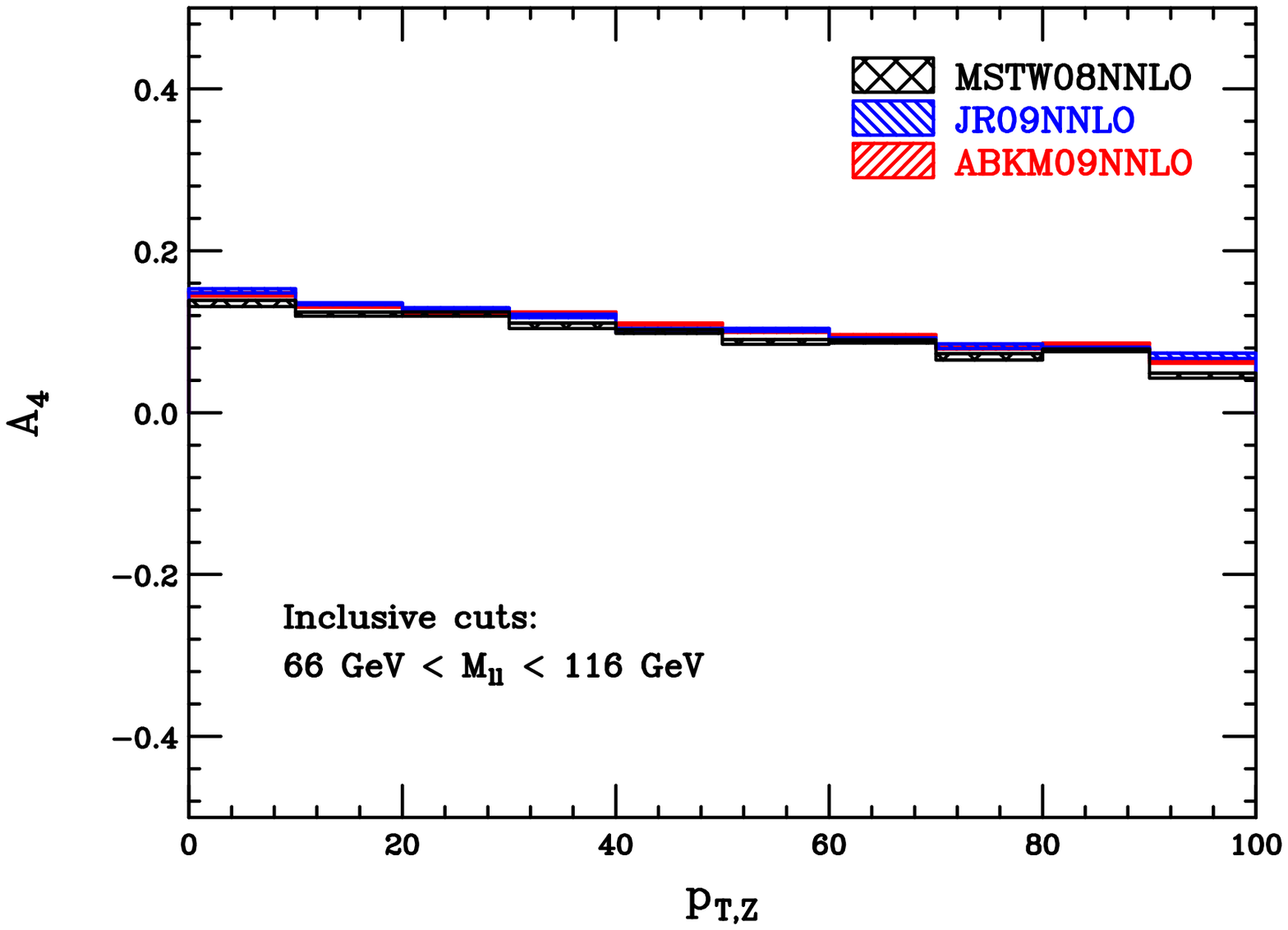}
\caption{Bin-integrated results for the CS moments $A_3$ (upper panel) and $A_4$ (lower panel), presented as a function of the lepton-pair transverse momentum, 
for all three NNLO PDF sets.  
Only a cut on the invariant mass $66 \, {\rm GeV} \leq M_{ll} \leq 116 \, {\rm GeV}$  has been implemented.  The bands indicate the PDF uncertainties for each set.  We note that the smallness of the PDF errors makes it difficult to distinguish 
the three separate bands in the plot.
\label{CS3}}
\end{center}
\end{figure}

Angular distributions in the Collins-Soper frame~\cite{Collins:1977iv} yield information on both the couplings of the $Z$-boson to leptons, and on the 
perturbative QCD which produces the $Z$ transverse momentum.  The differential cross section is expressed using the polar and azimuthal decay angles of the lepton in this frame 
as~\cite{Mirkes:1992hu,Mirkes:1994dp}
\begin{eqnarray}
\frac{d \sigma}{dM_{ll}^2 \,dp_{T,Z}^2 \,dY_Z \,d {\rm cos} \,\theta \,d \phi} &\sim& 1+{\rm cos^2}\,\theta + \frac{1}{2}A_0 \left(1-3 \,{\rm cos}^2 \theta \right)
+A_1 \,{\rm sin}\, 2\theta \, {\rm cos} \,\phi \nonumber \\
&+& \frac{1}{2}A_2 \, {\rm sin}^2\theta \, {\rm cos} \, 2\phi +A_3 \, {\rm sin} \,\theta\,{\rm cos}\, \phi+A_4 \, {\rm cos}\,\theta +...,
\end{eqnarray}
where the ellipses denote additional terms $A_{5,6,7}$ which we do not consider here.  We denote the $A_i$ as Collins-Soper (CS) moments.  The moments $A_3$ 
and $A_4$ are proportional to the parity-violating couplings of the quark and leptons.  $A_4$ is the only non-vanishing moment at LO.  All 
others are generated by additional radiation 
recoiling against the lepton pair.  These moments can be obtained using orthogonality relations for the trignometric functions they multiply.  
Defining the moment of a quantity $m$ as
\begin{equation}
\langle m \rangle = \frac{ \int d{\rm cos}\, \theta \, d\phi \, \;m \;d\sigma\left(M_{ll},p_{T,Z},Y_Z,{\rm cos}\,\theta,\phi\right)}
{ \int d{\rm cos}\, \theta \, d\phi \; d\sigma\left(M_{ll},p_{T,Z},Y_Z,,{\rm cos}\,\theta,\phi\right)},
\label{momdef}
\end{equation}
we can obtain the CS moments in the following fashion:
\begin{eqnarray}
\langle \frac{1}{2}(1-3\,{\rm cos}^2\theta )\rangle &=& \frac{3}{20} \left( A_0 - \frac{2}{3}\right), \nonumber \\
\langle {\rm sin}\, 2\theta \, {\rm cos}\, \phi \rangle = \frac{1}{5}A_1&,&
\langle {\rm sin}^2 \theta\, {\rm cos}\, 2\phi \rangle = \frac{1}{10}A_2,\nonumber \\
\langle {\rm sin}\, \theta\, {\rm cos}\, \phi \rangle = \frac{1}{4}A_3 &,&
\langle {\rm cos}\, \theta \rangle = \frac{1}{4}A_4.
\label{CSproj}
\end{eqnarray}
We begin by showing the results for the normalized ${\rm cos} \,\theta$ distribution in Fig.~\ref{CSctheta}.  The PDF errors are completely negligible 
for this distribution, and all three sets are in perfect agreement.  This prediction of perturbative QCD is stable against theoretical uncertainties.  
However, we will show later the important effect that acceptance cuts have on the ${\rm cos}\,\theta$ distribution.  We next display the results for the CS 
moments in Figs.~\ref{CS1},~\ref{CS2},~and~\ref{CS3}.  These again show very little variation under choice of PDF set or 
eigenvector within a set.  The CS 
moments are also affected dramatically by acceptance cuts, as we demonstrate later.

In order to reduce backgrounds affecting the Drell-Yan measurement, LHC experimentalists have discussed imposing a cut demanding that the two leptons not 
be back-to-back in the transverse plane~\cite{mschmitt}.  $Z$-boson events where the leptons are separated by nearly 180$^{\circ}$ suffer from large backgrounds from 
semi-leptonic $b$ decays.  A cut demanding a minimum value for this angle reduces such backgrounds.  To study the QCD predictions in the presence of this cut, we denote by $\Delta \phi_{ll}$ the lower cut on the separation angle on the two leptons.  For example, a $\Delta \phi_{ll}$ cut of $3^{\circ}$ denotes that all events where the deviation between the leptons in the transverse plane is more than $3^{\circ}$ are accepted.  Two possible problems make 
this cut worrisome from the perspective of stability under QCD corrections.  The region $\Delta \phi_{ll} \sim 0$ is dominated by the emission of soft and collinear 
gluons, and large logarithms of $\Delta \phi_{ll}$ invalidate fixed-order predictions for this quantity.  Resummation is required.  For $\Delta \phi_{ll} > 0$, the 
perturbative expansion starts at ${\cal O}(\alpha_s)$.  Our result is effectively only next-to-leading order for this observable, and could be obtained by studying $Z$+1 jet at NLO.  
This indicates that the scale variation will be greater than for the inclusive result.

\begin{figure}[h!]
\begin{center}
\includegraphics[width=4.75in]{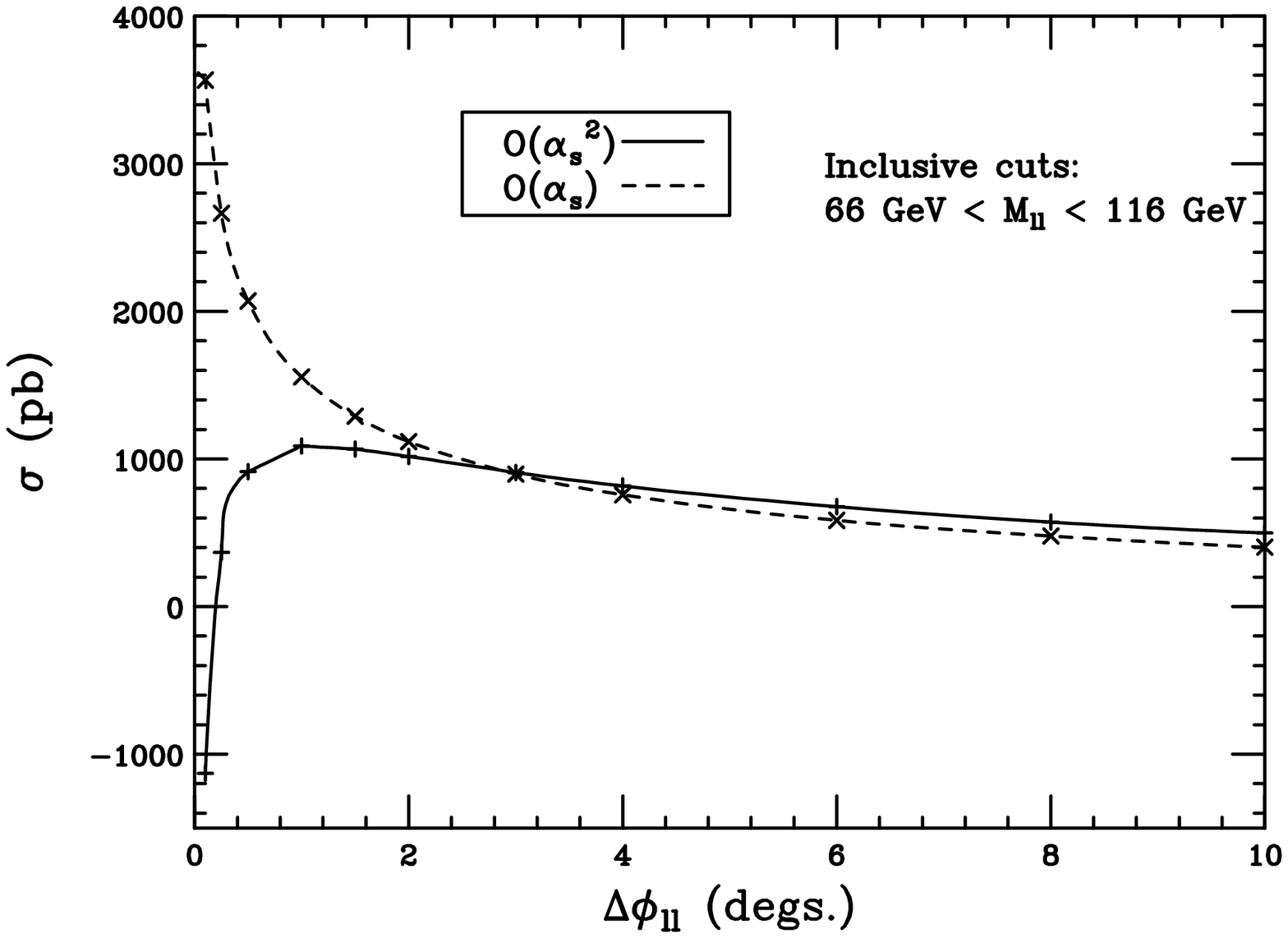}
\caption{Results obtained by including a lower cut on the transverse angular separation between the two leptons, $\Delta \phi_{ll}$, at ${\cal O}(\alpha_s)$ and ${\cal O}(\alpha_{s}^2)$.  Only a cut on the invariant mass $66 \, {\rm GeV} \leq M_{ll} \leq 116 \, {\rm GeV}$  has been implemented.
\label{dPhi}}
\end{center}
\end{figure}

To study these issues we show in Fig.~\ref{dPhi} the results obtained by including all events with angular separation greater than some lower cut $\Delta \phi_{ll}$ at ${\cal O}(\alpha_s)$ and ${\cal O}(\alpha_{s}^2)$; for the inclusive cross section these would respectively be the NLO and NNLO cross sections.  The onset of large logarithms as 
$\Delta \phi_{ll} \to 0$ is clearly visible in the divergence of the cross sections toward the left side of the plot.  In this region a resummation of the associated 
logarithms is required.  Fixed-order perturbation theory cannot be trusted below a cut value of roughly $\Delta \phi_{ll} \approx 3^{\circ}$, where the bands cross.  To study the 
theoretical uncertainty on the cross section and acceptance in the region where fixed-order results should give a reasonable estimate of the cross section, 
we show below the results including scale and PDF errors for the choice $\Delta \phi_{ll}=3^{\circ}$ and with MSTW PDFs:
\begin{eqnarray}
\sigma_{\Delta \phi=3^{\circ}} &=& 903.8^{+26.5}_{-24.1}{\rm (scale)}^{+28.0}_{-25.5}{\rm (PDF)}\; {\rm (pb)}; \nonumber \\
A_{\Delta \phi=3^{\circ}} &=& 0.943^{+0.024}_{-0.020}{\rm (scale)}^{+0.004}_{-0.006}{\rm (PDF)}.
\end{eqnarray}
While the relative PDF errors on the cross section and acceptance are the same as for the other cuts studied above, the scale errors on both reach $\pm 2.5\%$.  This is 
significantly larger than the scale dependence found for the inclusive result or for the standard acceptance cut defined previously, and occurs because 
this obervable begins at one order higher in perturbation theory.  Although the experimental errors might be reduced by imposing this cut, the increased 
theoretical uncertainty should be accounted for in analyses.

\begin{figure}[h!]
\begin{center}
\includegraphics[width=5.0in]{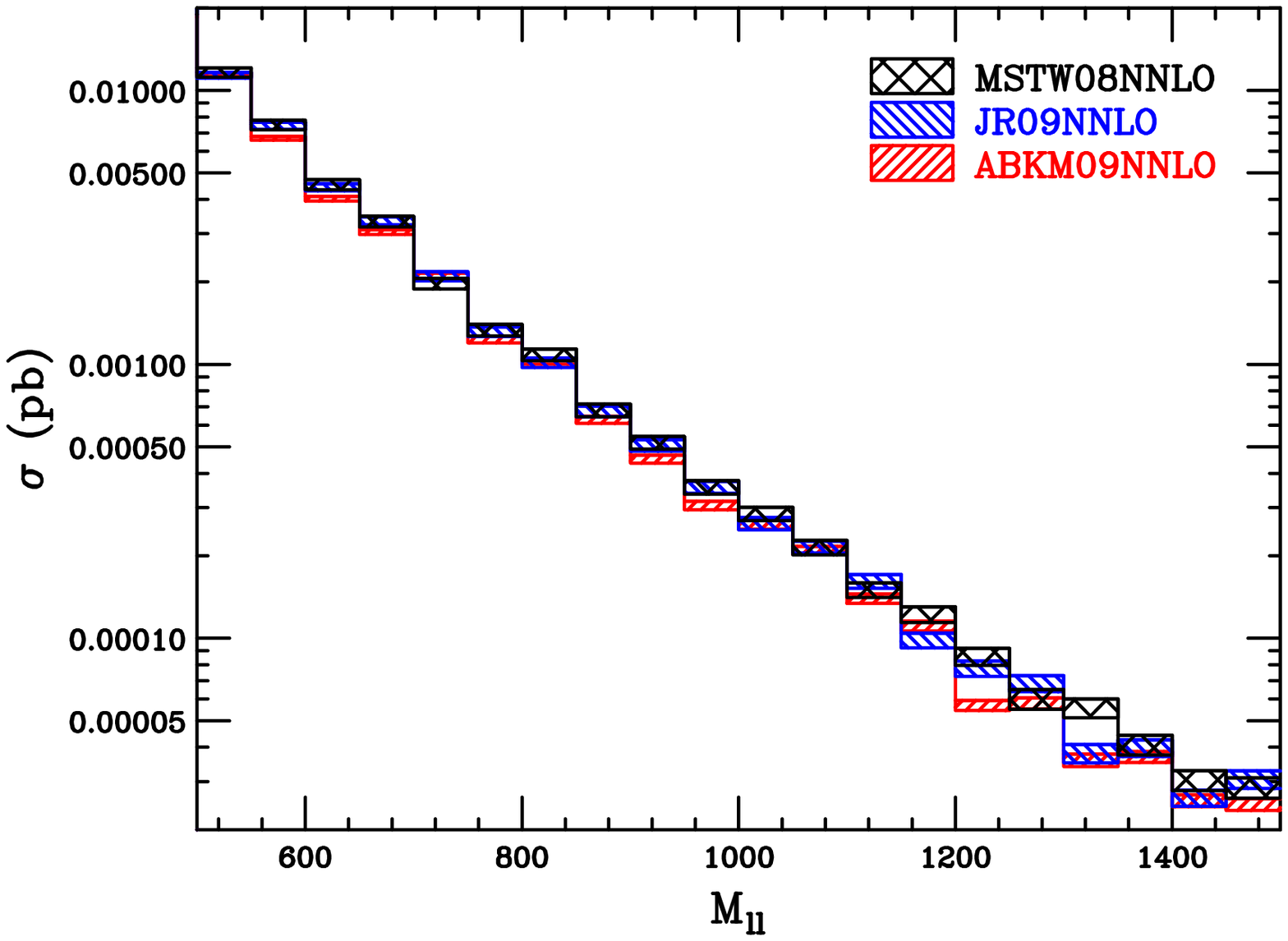}
\caption{Bin-integrated cross sections as a function of invariant mass for the range $500\, {\rm GeV} \leq M_{ll} \leq 1.5\, {\rm TeV}$.  No additional cuts have been imposed.
\label{minv}}
\end{center}
\end{figure}

As an additional phenomenological result relevant to LHC analyses, we study the invariant mass distribution far above the $Z$-pole.  Events in this phase-space region serve as a background to searches for high-mass resonances or contact interactions.  We run FEWZ once, and use the histogramming feature to study the high-mass portion of the $M_{ll}$ distribution.  The result for the range $500\, {\rm GeV} \leq M_{ll} \leq 1.5\, {\rm TeV}$ is shown in Fig.~\ref{minv}.  The JR 2009 PDF set gives consistently lower cross sections than both the MSTW and ABKM sets.  The MSTW and ABKM fits agree over the entire invariant mass range.  In this phase-space region, important additional corrections come from electroweak Sudakov logarithms.  These will be included in a future update of FEWZ.

\subsection{Results for distributions: {\it standard} cuts}

\begin{figure}[h!]
\begin{center}
\includegraphics[width=4.75in]{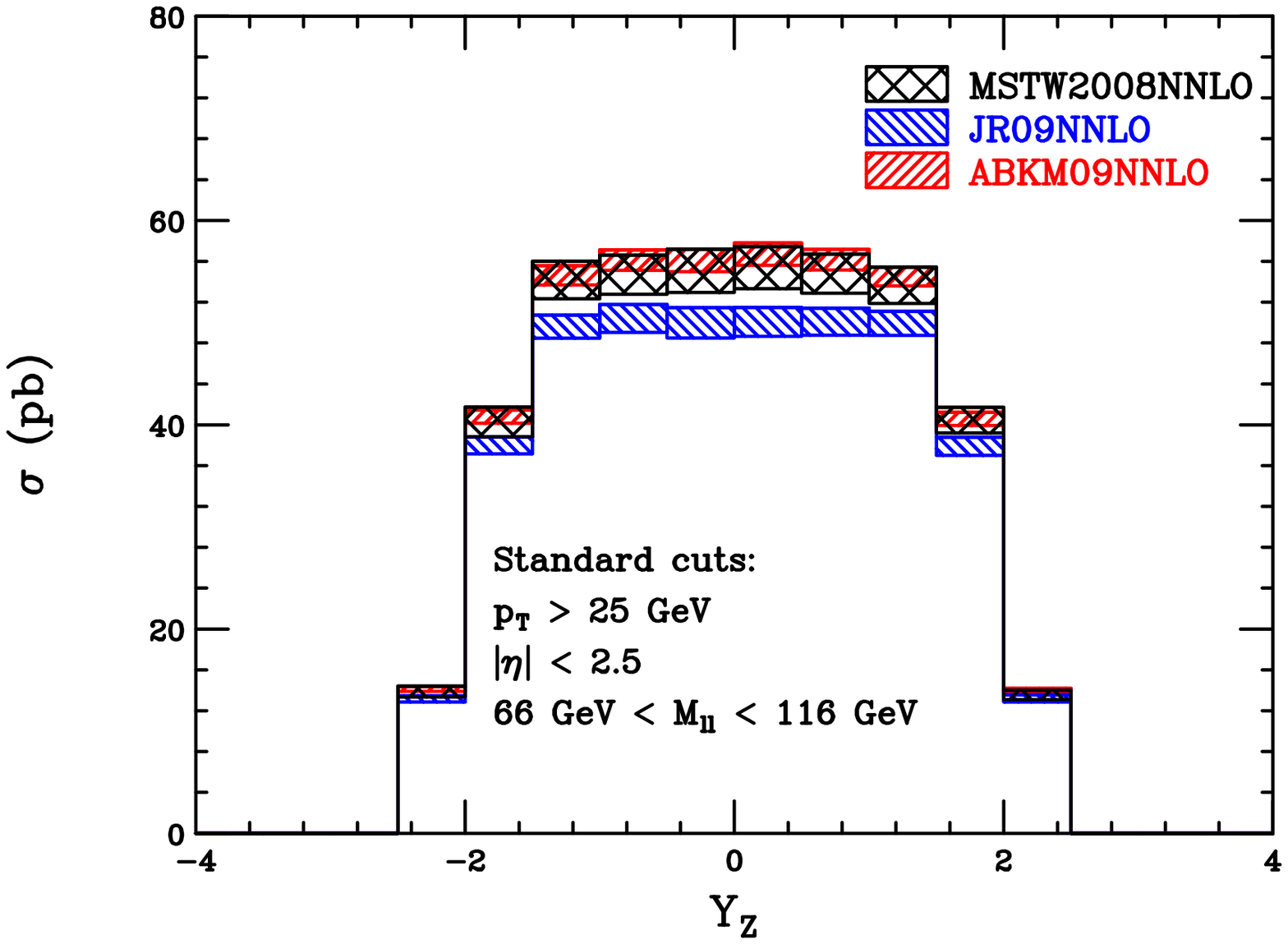}\\
\vspace{0.5cm}
\includegraphics[width=4.75in]{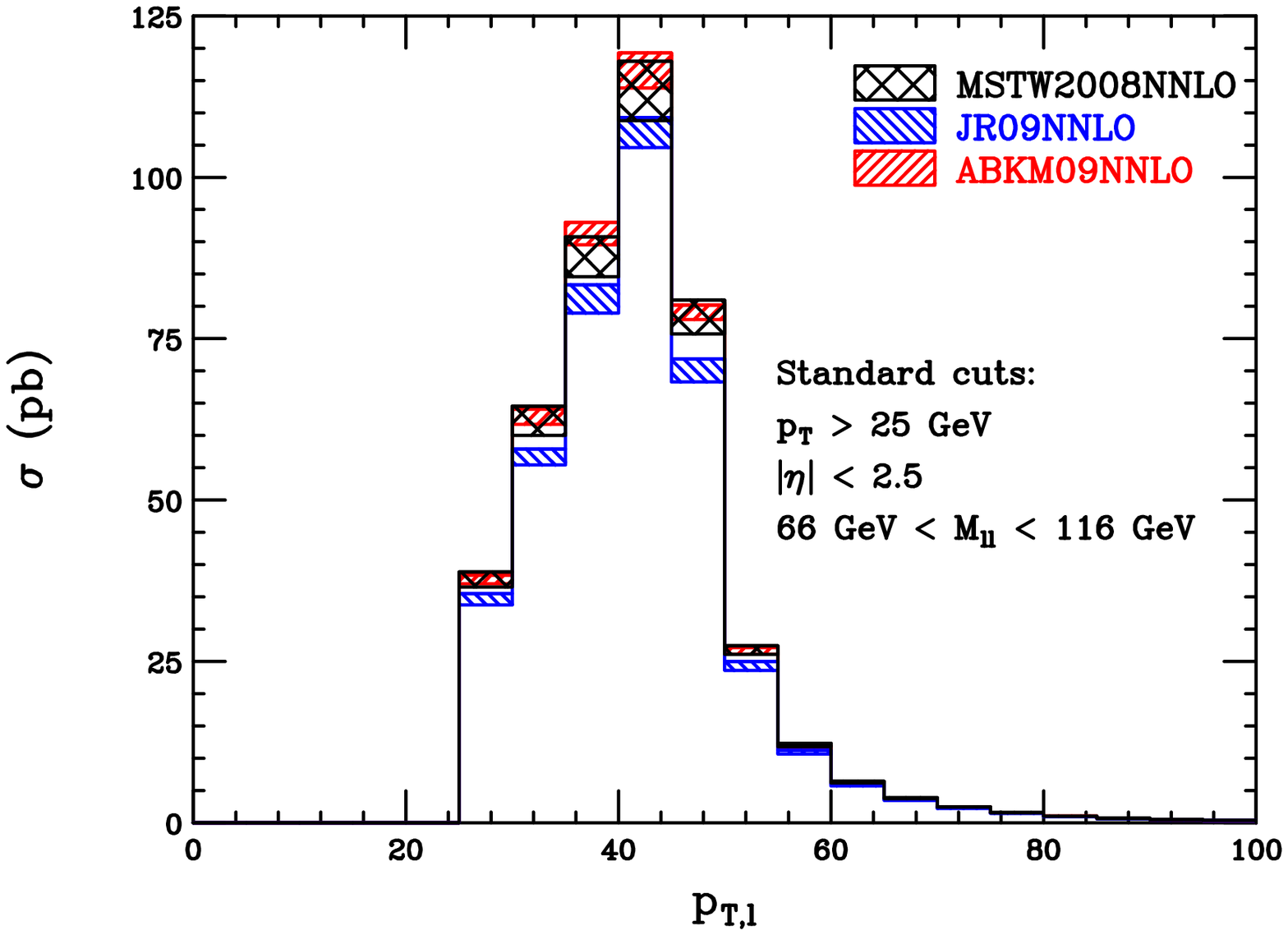}
\caption{Bin-integrated cross sections for the $Z$-boson rapidity (upper panel) and lepton transverse momentum (lower panel) for all three NNLO PDF sets.  
The standard acceptance cuts of Eq.~(\ref{standardacc}) have been implemented.  The bands indicate the PDF uncertainties for each set.
\label{stdbasic}}
\end{center}
\end{figure}

To demonstrate the effects of leptonic cuts on the studied distributions, we run FEWZ again for each of the three NNLO PDF sets and impose the {\it standard} 
acceptance cuts introduced previously:
\begin{eqnarray}
p_{T,lep} &>& 25 \,\, \text{GeV},\;\;\; |\eta_{lep}| < 2.5, \nonumber \\
\Delta R_{lep,lep} &>& 0.5,\;\;\; \Delta R_{lep,jet}>0.5.
\end{eqnarray}
We show only a few representative distributions to avoid too severe a proliferation of plots.  For basic kinematic distributions such as the transverse 
momenta or rapidities 
of the leptons or $Z$-boson, the standard acceptance cuts do not dramatically affect the shapes.  This is demonstrated for the lepton $p_T$ and $Z$-boson rapidity 
distributions in Fig.~\ref{stdbasic}.  The MSTW and ABKM $Z$-boson rapidity distributions become flatter after the standard acceptance cuts are 
implemented, suggesting that their differences from the JR set occur at low-$x$.  No other significant differences from the distributions obtained 
with only the invariant mass cut are apparent, besides the obvious changes in distribution endpoints.

\begin{figure}[h!]
\begin{center}
\includegraphics[width=4.5in]{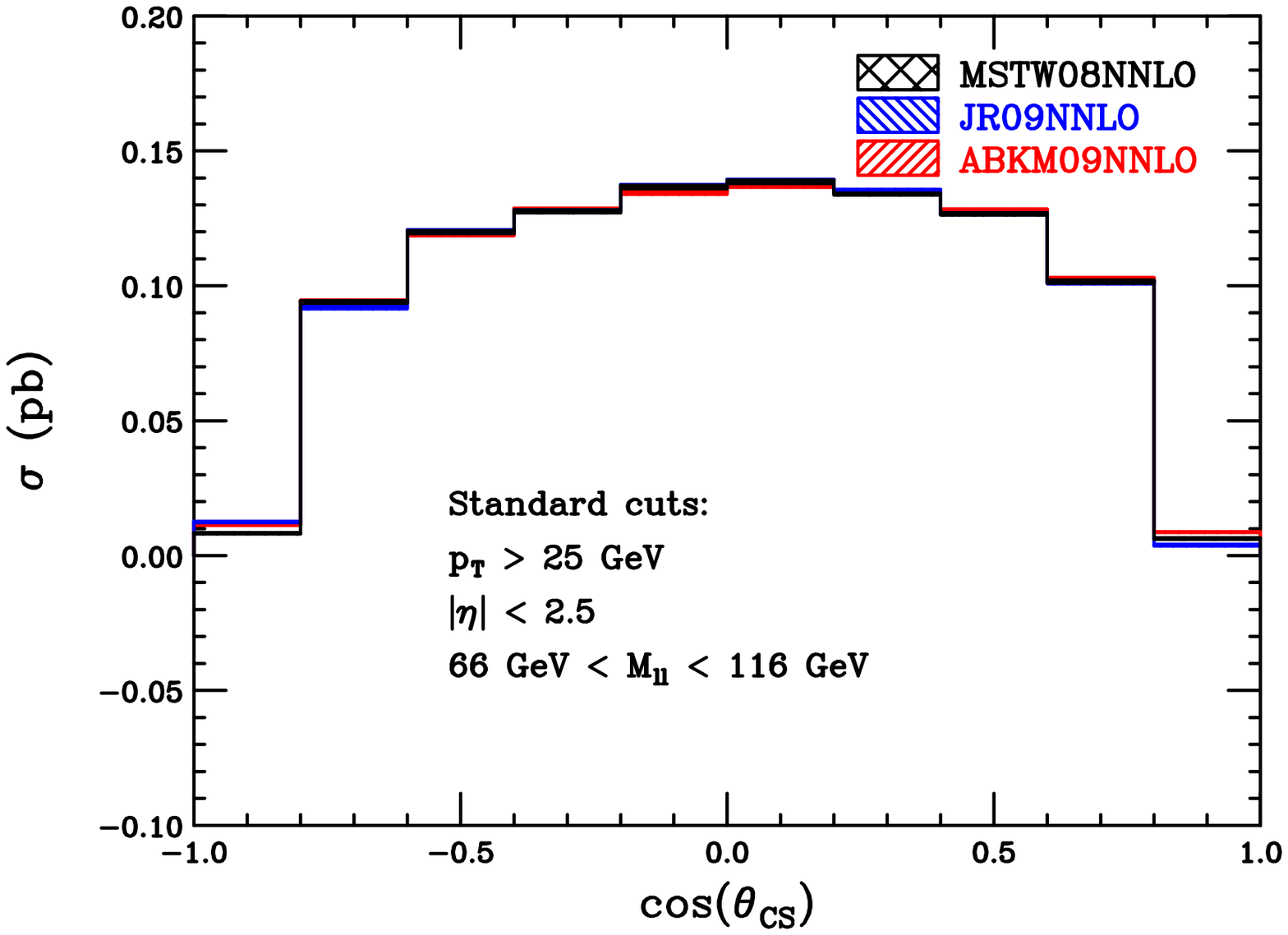}
\caption{Bin-integrated, normalized ${\rm cos}\,\theta$ distribution for all three NNLO PDF sets.  The polar angle of the lepton is defined in the 
Collins-Soper frame, as indicated by the subscript in the plot. The standard acceptance cuts of Eq.~(\ref{standardacc}) have been implemented.  
The bands indicate the PDF uncertainties for each set.  We note that the smallness of the PDF errors makes it difficult to distinguish 
the three separate bands in the plot.
\label{CScthetastd}}
\end{center}
\end{figure}

The situation changes drastically when angular distributions of the leptons are studied.  We begin by showing the ${\rm cos}\,\theta$ 
distribution after standard acceptance 
cuts are imposed in Fig.~\ref{CScthetastd}.  The shape of the distribution in Fig.~\ref{CSctheta} has completely changed upon addition of the cuts.  This can 
be understood by considering the leading-order kinematics.  At LO, the expressions for the lepton and anti-lepton pseudorapidities are given by
\begin{equation}
\eta_l = \frac{1}{2} \,{\rm ln}\left[ \frac{x_1}{x_2}\frac{1+{\rm cos} \,\theta}{1-{\rm cos} \,\theta}\right],\;\; 
\eta_{\bar{l}} = \frac{1}{2} \,{\rm ln} \left[\frac{x_1}{x_2}\frac{1-{\rm cos} \,\theta}{1+{\rm cos} \,\theta}\right].
\end{equation}
Events near ${\rm cos} \,\theta \approx \pm 1$ correspond to events with high values of $|\eta_{l,\bar{l}}|$, which are removed by the cuts in 
Eq.~(\ref{standardacc}).

\begin{figure}[h!]
\begin{center}
\includegraphics[width=4.5in]{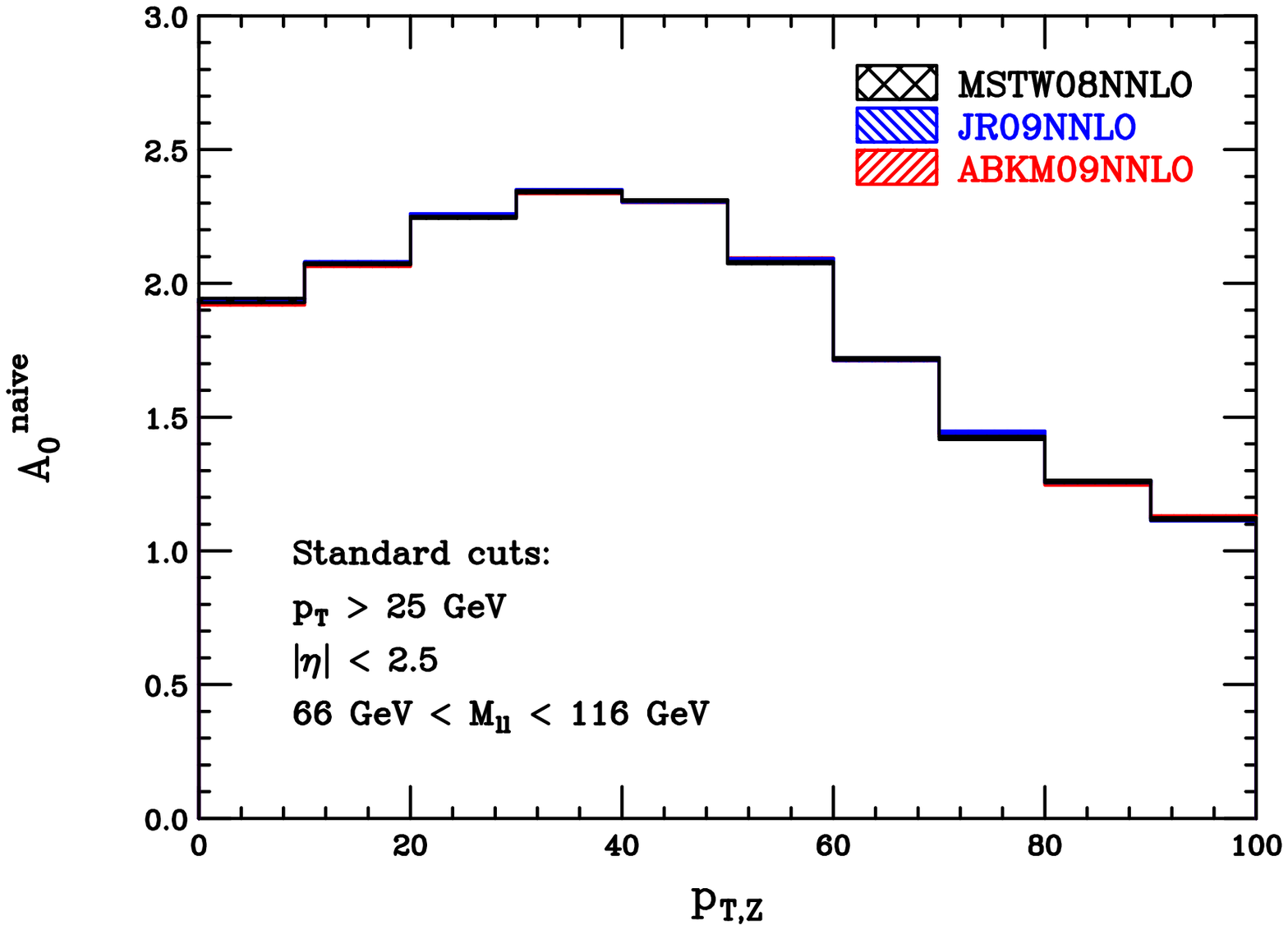}
\caption{Bin-integrated results for the naive CS moment $A_0^{naive}$, presented as a function of the lepton-pair transverse momentum, 
for all three NNLO PDF sets.  The standard acceptance cuts of Eq.~(\ref{standardacc}) have been implemented.  
The bands indicate the PDF uncertainties for each set.  We note that the smallness of the PDF errors makes it difficult to distinguish 
the three separate bands in the plot.
\label{A0std}}
\end{center}
\end{figure}

The modification of the polar angle distribution, and additionally the lower cut on the leptonic transverse momenta, have a dramatic effect 
on the determination of the CS moments defined previously.  The orthogonality conditions 
of Eq.~(\ref{CSproj}) no longer apply when only a finite region of ${\rm cos}\,\theta$ is integrated over experimentally.  To demonstrate the impact of 
these acceptance cuts on the extraction of the CS moments, we define `naive' CS moments $A_i^{naive}$ which we obtain by integrating 
over only the region allowed by the standard acceptance cuts in the moment definition of 
Eq.~(\ref{momdef}) .  The result for $A_0^{naive}$ is shown in Fig.~\ref{A0std}.  Instead of beginning at zero for 
$p_{T,Z}=0$ GeV and monotonically approaching $A_0 \approx 0.75$ at $p_{T,Z}=100$ GeV, $A_0^{naive}$ begins at $A_0^{naive} \approx 2$, rises to a 
maximum $p_{T,Z} \approx 45$ GeV, and falls to $A_0^{naive} \approx 1$ at $p_{T,Z}=100$ GeV.  These general features can be confirmed by analytic 
integration of the LO result for the cross section.  The acceptance cuts completely change the qualitative features of this distribution. A similar effect 
is obtained for the moment $A_2$, shown below in Fig.~\ref{A2std}.  Instead of a distribution monotonically increasing toward 0.75 at large $p_{T,Z}$ as seen in 
Fig.~\ref{CS2}, it instead monotonically descreases from zero toward $A_2^{naive}\approx -3$ in the presence of standard acceptance cuts.
\begin{figure}[h!]
\begin{center}
\includegraphics[width=4.5in]{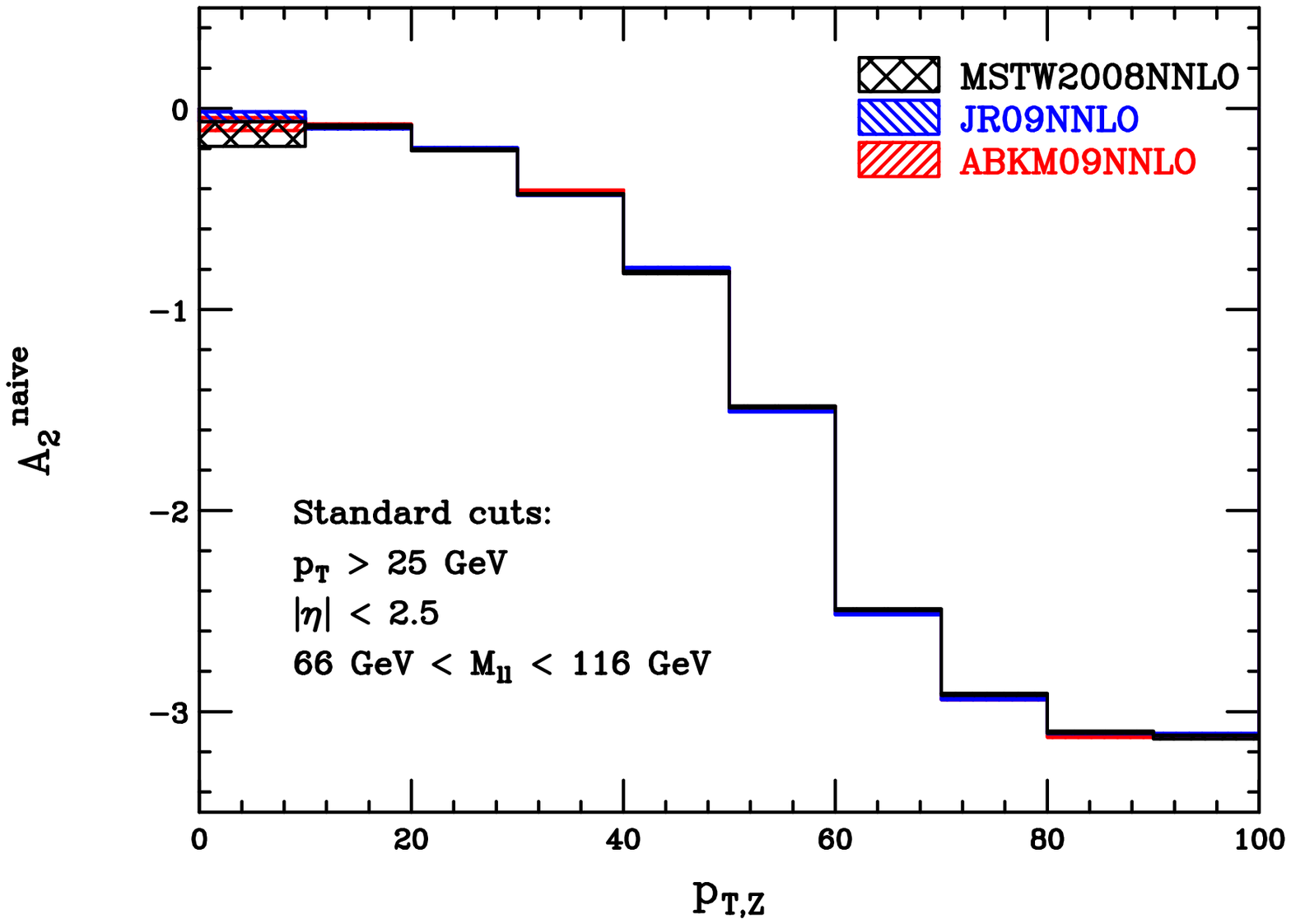}
\caption{Bin-integrated results for the naive CS moment $A_2^{naive}$, presented as a function of the lepton-pair transverse momentum, 
for all three NNLO PDF sets.  The standard acceptance cuts of Eq.~(\ref{standardacc}) have been implemented.  
The bands indicate the PDF uncertainties for each set.  We note that the smallness of the PDF errors makes it difficult to distinguish 
the three separate bands in the plot.
\label{A2std}}
\end{center}
\end{figure}

\section{Conclusions \label{sec:conc}}

In this manuscript we have presented an improved version of the analysis code FEWZ for the study of differential lepton-pair production through next-to-next-to leading 
order in perturbative QCD.  FEWZ allows the kinematics of the leptons and all associated hadronic radiation to be studied, and permits analysis of events both 
on and off the $Z$-peak.  The new features of the code include an efficient, parallelized integration routine that can take advantage of distributed computing 
systems such as {\it Condor}.  Sub-1\% technical precisions are easily obtainable even in the presence of severe phase-space cuts that accept only a small 
fraction of the inclusive cross section.  Histograms of most interesting kinematic variables are now filled during a single run of FEWZ, and 
PDF errors are automatically computed for each histogram bin.  Both inclusive results and distributions are obtained with a single run of FEWZ on a single multi-core 
desktop after no more than several days even with extreme cuts imposed, or in less time using a {\it Condor} system.

We have presented numerous phenomenological results relevant for LHC studies that also demonstrate the new features of FEWZ.  We have shown
inclusive cross section central values, scale variations and PDF errors for all three NNLO PDF fits, and have compared them 
to recent measurements by the ATLAS and CMS collaborations.  Additional predictions for 
representative acceptance cuts are also shown.  FEWZ delivers a wealth of additional information in the form of differential distributions. We have presented and 
discussed several basic lepton kinematic distributions.

The most interesting phenomenological results we have obtained are associated with the angular distributions of the leptons.  We have studied a proposed 
cut on the transverse-plane angular separation between the leptons, which is designed to reduce backgrounds in precision studies of the $Z$-peak.  This distribution 
suffers from large logarithms if the leptons are nearly back-to-back, rendering fixed-order perturbation theory unstable.  In the region where fixed-order 
calculations can be trusted, imposing this cut increases the scale uncertainty from sub-1\% to $\pm 2.5\%$, since the prediction begins at ${\cal O}(\alpha_s)$ 
for this observable.  The increased theoretical uncertainty must be accounted for in precision studies.  We have additionally studied the angular distributions 
of the leptons in the Collins-Soper frame, and the associated Collins-Soper moments.  These quantities are absolutely stable under perturbative corrections, 
but the effect of acceptance cuts are dramatic, and change their qualitative features.

FEWZ is intended for use in studies of all aspects of lepton-pair production at hadron colliders where fixed-order perturbation theory is applicable: for computing 
inclusive cross sections, distributions of basic kinematics and of angular quantities, and especially when determined the effects of limited detector acceptance.  
In order to compare experimental results with the most precise available theory, NNLO QCD should be utilized in all stages of the analysis, including when acceptance 
corrections are determined.  FEWZ makes this possible.  It is a flexible framework that makes future inclusion of electroweak corrections and other 
improvements simple.  We look forward to its continued use in experimental studies.

\bigskip
\bigskip
\noindent
{\bf{\Large Acknowledgements}}
\bigskip

\noindent
We are grateful to F.~Stoeckli for inspiring and advising us on the inclusion of the histogramming feature in FEWZ, and on the restructuring 
of the numerical integration.  We thank S.~Yost and J. Qian for 
valuable feedback on the original version of FEWZ, and M. Schmitt for useful discussions on experimental capabilities and desires.  We also 
thank W.~Sakumoto for alerting us to parity-violating moments for inclusion in our code, T. Hahn for feedback regarding CUBA, and K. Melnikov for helpful comments.  
Work supported in part by the US Department of Energy, Division of High Energy Physics, under Contract DE-AC02-06CH11357.

\end{document}